% ...

\documentclass[preprint,12pt]{elsarticle}

\usepackage{hyperref}

\usepackage{multirow}
\usepackage[table,xcdraw]{xcolor}
\usepackage{tabularx,booktabs}
\usepackage{array}

\usepackage{graphicx} 	  
\usepackage{amssymb} 	

\usepackage{color}
\usepackage{amsmath}
\usepackage{mathtools}
\usepackage{fullpage}
\usepackage{algorithmic}
\DeclareMathOperator*{\argmin}{argmin}
\DeclareMathOperator*{\argmax}{argmax}
\algsetup{linenosize=\small}
\usepackage[table,xcdraw]{xcolor}
\usepackage{multirow}
\usepackage[super]{nth}
\usepackage{graphicx,adjustbox}
\usepackage{caption}
\usepackage[labelformat=simple]{subcaption}
\usepackage{comment}
\usepackage{setspace}
\usepackage{textcomp}
\usepackage{xspace}
\usepackage{siunitx}
\usepackage{epsfig}
\usepackage{epstopdf}
\usepackage{soul}
\usepackage{url}
\usepackage{tablefootnote}
\usepackage[linesnumbered,ruled]{algorithm2e}
\usepackage{booktabs}
\usepackage{hyperref}
\usepackage[normalem]{ulem}
\usepackage{footnote}
\usepackage[misc,geometry]{ifsym} 
\makesavenoteenv{tabular}

\let\OldTexttrademark\texttrademark
\renewcommand{\texttrademark}{\OldTexttrademark\xspace}%

\usepackage{epsfig}
\usepackage{amssymb}

\usepackage{tikz}
\usepackage{tkz-tab}
\usetikzlibrary{automata,arrows,positioning,calc}
\usetikzlibrary{shapes,snakes}

\def\Plus{\texttt{+}}

\DeclarePairedDelimiter\abs{\lvert}{\rvert}%
\DeclarePairedDelimiter\ceil{\lceil}{\rceil}

\newcolumntype{C}{>{\centering\arraybackslash}X} 
\newcolumntype{b}{X}
\newcolumntype{s}{>{\hsize=.5\hsize}X}
\newcolumntype{v}{>{\hsize=.3\hsize}X}

\newcommand{\newver}[1]{{#1}}
\journal{Computer Networks}

\begin{document}
	
	\begin{frontmatter}
		
	\title{To Overlap or not to Overlap: Enabling Channel Bonding in \newver{High-Density} WLANs}
	
	\author{Sergio~Barrachina-Mu\~noz \corref{cor1}} 
	\author{Francesc Wilhelmi}
	\author{Boris~Bellalta}
	\address{Wireless Networking (WN), Universitat Pompeu Fabra, Barcelona, Spain}
	\cortext[cor1]{Corresponding Author: Sergio Barrachina-Mu\~noz, Email address: \href{sergio.barrachina@upf.edu}{sergio.barrachina@upf.edu}}
	
	\begin{abstract}
		
		Wireless local area networks (WLANs) are the most popular kind of wireless Internet connection \newver{because of their simplicity of deployment and operation. As a result}, the number of devices accessing the Internet through WLANs such as laptops, smartphones, or wearables, is increasing drastically at the same time that applications' throughput requirements do. 
		To cope with these challenges, channel bonding (CB) techniques are used for enabling higher data rates by transmitting in wider channels\newver{, thus increasing spectrum efficiency}. However, \newver{important issues like} higher potential co-channel and adjacent channel interference arise when bonding channels.
		This may harm the performance of the carrier sense multiple access (CSMA) protocol \newver{because of recurrent backoff freezing, while making nodes more sensitive to hidden node effects}.
		In this paper, we address \newver{the following} point at issue: is it convenient for \newver{high-density (HD)} WLANs to use wider channels and potentially overlap in \newver{the} spectrum? \newver{First, we highlight key aspects of DCB in toy scenarios through a continuous time Markov network (CTMN) model. Then,} by means of extensive simulations \newver{covering a wide range of traffic loads and access point (AP) densities}, we show that dynamic channel bonding (DCB) -- which adapts the channel bandwidth on a per-packet transmission -- significantly outperforms traditional single-channel \newver{on average}. \newver{Nevertheless, results also corroborate that DCB is more prone to generate unfair situations where WLANs may starve. Contrary} to most of the current thoughts \newver{pushing} towards non-overlapping channels in HD deployments, we highlight the benefits of allocating channels as wider as possible to WLANs altogether with implementing \newver{adaptive} access policies to cope with the unfairness situations that may appear.
		
	\end{abstract}

	\begin{keyword}
		dynamic channel bonding, WLAN, spatial distribution, traffic load, IEEE 802.11ax	
	\end{keyword}
	
	\end{frontmatter}
	

	\section{Introduction}\label{sec:introduction}	
	
	Although remarkable technological improvements have been achieved in the last decades, wireless local area networks (WLANs), with IEEE 802.11's Wi-Fi as the most widely used standard, still face important challenges that degrade their performance. \newver{Particularly, the} frequency spectrum is becoming scarce and inefficient because of the rising number of wireless devices, the characteristically heterogeneous and random WLAN deployments, and the raising throughput demands (e.g., some virtual reality applications require more than 1 Gbps to operate properly \cite{elbamby2018toward}). All these circumstances lead to dense \newver{or high-dense (HD)} scenarios with coexistence issues \newver{since} WLANs \newver{try} to selfishly serve their users in non-collaborative deployments.
	
	As a result, there is a clear need of exploiting the spectrum in a more efficient way by maximizing transmissions' bandwidth. One of the most promising techniques to overcome such a challenge is channel bonding (CB). The main idea behind CB is to allow using wider \newver{bandwidths} in order to transmit at higher transmission rates, increasing the \newver{potential} throughput accordingly.\footnote{\newver{According to the well-known Shannon-Hartley capacity theorem, the capacity (or raw throughput) of a channel increases with the bandwidth.}} CB for WLANs was firstly introduced in the IEEE 802.11n-2009 amendment \cite{80211n} by letting two separated 20 MHz channels (or basic channels) get combined into a 40 MHz channel. Later, IEEE 802.11ac-2013 \cite{80211ac} introduced the capability of transmitting also in 80 and 160 MHz channels. Future amendments like the IEEE 802.11ax-2019 \cite{802.11ax} or \newver{EXtreme throughput (XT) \cite{80211ET}, which is expected to support up to 320 MHz transmissions, will boost the use of wider channels}. A survey of CB schemes for different types of wireless networks is provided in \cite{bukhari2016survey}.
	
	There are important drawbacks, however, when it comes to transmitting in wider channels: essentially, the larger the bandwidth used for transmitting, the wider the \newver{spectrum} suffering from co-channel and adjacent channel interference at neighboring nodes. That is, CB may be counterproductive since WLANs nearby are more likely to partially overlap, causing severe performance degradation due to the listen-before-talk nature of the \newver{carrier sense multiple access (CSMA)} protocol. This effect is further exacerbated when following static channel bonding (SCB) \cite{park2011ieee}. \newver{Besides, the signal-to-interference-plus-noise ratio (SINR) at the receiver decreases for wider channels since the transmission power is spread through the whole transmission bandwidth (or subcarriers). Accordingly, situations like the hidden node problem are more prone to occur when implementing CB \cite{deek2011impact}.}
	
	In this regard, dynamic channel bonding (DCB) allows adapting the selected transmission bandwidth to the channel status right before transmitting. This provides a higher degree of flexibility that improves the instantaneous throughput in a simple and efficient way.
	Then, we can differentiate two approaches with respect to spectrum management in WLANs: \textit{i}) fostering transmissions in non-overlapping basic channels, or \textit{ii}) enabling faster transmissions in wider channels that may potentially overlap in \newver{the} spectrum. Alas, in high density (HD) spatially distributed scenarios,\footnote{In spatially distributed scenarios, nodes are not \newver{necessarily} within the carrier sense range of each other. Thus, different groups of potentially overlapping WLANs may be given.} the complex interrelations among nodes (located inside or outside the carrier sense range of each other) complicate the task of \textit{a priori} estimating the optimal spectrum management approach on a per-WLAN basis.
	
	
	Significant research has been conducted on the impact of DCB on spatially distributed WLANs’ performance under saturation regimes. However, to the best of our knowledge, the effects of unsaturated traffic patters, which fit better to real world problems, are still unknown. While saturated regimes offer valuable insights on \newver{worst-case} scenarios, WLANs are characteristically unsaturated with load patterns that deeply depend on the application/s being supported. In such scenarios, overlapping approaches seem to be even more convenient since the sensed channels usually remain free during larger periods of time.
	
	In this paper\newver{,} we compare the performance of traditional single-channel with channel bonding (including a stochastic \newver{variant}) in networks under different traffic load regimes. To do so, we first introduce an analytical model to depict the behavior of the aforementioned CB approaches (or policies) in spatially distributed WLAN networks. The model is based on continuous time Markov networks (CTMNs) and captures \newver{both saturated and unsaturated} regimes. Then, by means of simulations, we evaluate the performance of the CB policies in terms of throughput and delay in toy scenarios and HD WLAN deployments. We find that for low \newver{individual and} neighboring traffic loads single-channel can improve CB in terms of delay since the time to access the channel is reduced. However, in general, DCB significantly outperforms traditional single-channel in most of the evaluated scenarios. \newver{Nonetheless, DCB is more prone to cause starvation and hidden nodes, which may lead to unfair scenarios with highly unbalanced performance among WLANs}. Results suggest that future WLANs should be allocated all the available bandwidth and dynamically adapt to the spectrum.
	
	The remainder of this article is organized as follows. In Section \ref{sec:dcb}, we introduce CB for IEEE 802.11 WLANs, \newver{present the related work}, and define the policies considered in this work. Then, in Section \ref{sec:use_cases} we analytically model the interactions of DCB in spatially distributed deployments and \newver{point out key aspects of} its performance through toy scenarios. DCB in HD WLANs is assessed in Section \ref{sec:evaluation} \newver{by means of simulations}. We conclude with some final remarks at Section \ref{sec:conclusions}.
	
	\section{Channel bonding} \label{sec:dcb}
	
	\subsection{Channel bonding in IEEE 802.11 WLANs}
	
	CB is a technique whereby nodes, i.e., access points (APs) and stations (STAs), are allowed to use contiguous sets of available basic channels for their transmissions, thus potentially achieving higher throughput. Namely, by doubling the channel bandwidth, approximately the double data capacity can be achieved \newver{if the modulation coding scheme (MCS) is kept}.
	CB for WLANs was firstly introduced in the IEEE 802.11n-2009 amendment \cite{80211n}, where high throughput (HT) STAs are allowed to transmit in more than one 20 MHz channel (or basic channel). Specifically, this amendment allowed bonding up to two basic channels composing a 40 MHz channel in the 2.4/5 GHz bands. Works in the literature \cite{arslan2010auto,deek2011impact, arslan2013acorn} show important improvements achieved with CB in IEEE 802.11ac networks when properly adjusting the transmission power and data rates in WLANs operating at 5 GHz. Note that in the traditional 2.4 GHz band, CB has been found to be counterproductive since only three non-overlapping basic channels are allowed \cite{shrivastava2008802}.
	
	Later, the IEEE 802.11ac-2013 amendment \cite{80211ac} increased the maximum number of bonded 20 MHz channels to 8, allowing very high throughput (VHT) STAs to transmit in up to 160 MHz in the 5 GHz band. Currently, the IEEE Task Group 11ax (TGax) is working on the IEEE 802.11ax amendment \cite{802.11ax}, which is expected to be published by 2019. As in IEEE 802.11ac, high efficiency (HE) STAs are also allowed to bond up to 8 basic channels. \newver{Moreover, the frequency spectrum efficiency is expected to be boosted in IEEE 802.11ax by combining orthogonal frequency-division multiple access (OFDMA) \cite{bellalta2016ieee} with preamble puncturing, an optional capability for enabling non-contiguous CB}. Recently, the EXtreme throughput (XT) study group \cite{80211ET} has been created with the objective of increasing the peak throughput and capacity of WLANs. The motivation behind XT are the expectations that more than 1 GHz of additional unlicensed spectrum may be available around 2020. \newver{Thus}, it will allow exploiting further spectrum at the 6 GHz band \newver{by transmitting in bandwidths up to 320 MHz}. \newver{Note that in this manuscript we consider only contiguous bandwidths. That is, we use representative values of IEEE 802.11ax (e.g., 78.125 kHz subcarrier spacing or up to 1024-QAM MCS) to provide results in an appropriate scale when assessing its potential on contiguous DCB. Non-contiguous CB is left for future works, that can take the results presented in this paper as a baseline.}

	Notwithstanding, implementing CB in ever-increasingly complex WLAN \newver{deployments} requires a careful balance of trade-offs.
	First, regarding channelization,\footnote{Channelization is the process of setting independent channels on neighboring APs in order to avoid interference among their WLANs.} the density of neighboring nodes and the number of independent basic channels (which are regulated by governmental institutions) determine the feasibility of deploying interference-free networks. Essentially, as transmission channels get wider, frequency spectrum reuse becomes arduous, and the probability of packet collisions due to co-channel and adjacent channel interference increases.
	Secondly, the higher the bandwidth, the smaller the transmitted power per Hz and corresponding coverage range. This, on the one hand, reduces the interference with other WLANs operating in a (partially) overlapping spectrum. On the other hand, it reduces the SINR at the destination STAs, resulting in lower transmission rates if the receiver is not close enough to the transmitter. \newver{In addition, higher packet loss rates can arise because of the hidden node problem resulting from the lower SINR at higher bandwidths.}
	\newver{In this regard, authors in \cite{deek2014intelligent} show that parameters like the strength of neighboring links and interferer loads strongly affect the performance of CB}.
	
	In \newver{short}, the multiple spatial distribution factors such as transmission powers, clear channel assessment (CCA) levels, allocated channels, or environment's path loss, make it really difficult to generalize to an optimal set of rules for transmission channel selection. It follows that bandwidth adaptation is required in order to cope with the challenging scenarios of next-generation WLANs.
	
	\subsection{\newver{Related work}}

	\newver{Since its emergence in the IEEE 802.11n amendment, CB has shown a great potential in WLANs. In \cite{arslan2010auto, deek2011impact,arslan2013acorn}, authors provides insight into the factors affecting CB performance in IEEE 802.11ac WLANs. These works show that increasing the bandwidth incurs into a lower SINR at the receivers, thus compromising its effectiveness. The fact is that transmitting in larger channel widths entails a reduction of Watt/Hertz, which accentuates the vulnerability to interference. In order to palliate the inherent limitations, several solutions have been proposed for 802.11 WLANs. Authors in \cite{deek2013joint} propose ARAMIS, a CB solution for simultaneously adapting the rate and channel bandwidth, boosted by spatial diversity in multiple-input multiple-output (MIMO) IEEE 802.11ac WLANs. Alternatively, \cite{wang2016managing} formulates a distributed CB scheme based on adaptive channel clear assessment (CCA).}
	
	\newver{Newer amendments like IEEE 802.11ac or IEEE 802.11ax broaden the capabilities of CB by providing larger channel widths (up to 160 MHz), thus accentuating the advantages and drawbacks of potential spectrum access solutions. Nonetheless, it is worth noting that next-generation deployments will be characterized by short-range WLAN scenarios \cite{bellalta2016ieee}. This reduction of the AP-STA distance, altogether with the usage of techniques like spatial diversity MIMO\cite{deek2013joint}, contributes to palliate issues regarding low SINR values. There are several works on CB for IEEE 802.11ac relying on simulations. For instance, \cite{gong2011channel, park2011ieee,barrachina2018performance} show significant throughput gains compared to single-channel. Authors in \cite{zeng2014first} conduct an empirical study corroborating the performance gains of CB in IEEE 802.11ac WLANs. However, these works also highlight that such gains are importantly affected by the operation of neighboring networks. To the best of our knowledge, there are not experimental works yet on CB for preliminary implementations of IEEE 802.11ax.}
	
	\newver{One of the first fine-grained spectrum access design to dynamically change both the channel width and center frequency was formulated in \cite{yun2013fine}. This approach, compatible with IEEE 802.11a WLANs, was proven to significantly outperform static allocations. More recently, a frame-level wideband spectrum adaptation prototype supported by specially-constructed preambles and spectrum detection is presented in \cite{wang2017wideband}. In this regard, a collision detection protocol for DCB is presented in \cite{huang2016dynamic}. Authors in \cite{chen2018dbs} designed a dynamic bandwidth selection protocol to diminish the \textit{carrier sensing decreasing} and \textit{outside warning range problems}. Other works proposed a heuristic primary channel selection for CB users \cite{khairy2018renewal},  a probabilistic spectrum distribution framework considering uncertain traffic load \cite{nabil2017adaptive}, or a prototype implementation for IEEE 802.11ac based on time-domain interference cancellation \cite{byeon2018reconn}.}
	
	\newver{Analytical models have been also widely used in the literature on CB in WLANs\cite{bellalta2014channel, bellalta2016interactions, han2016performance, kim2017throughput, khairy2018renewal}. CB in short-range IEEE 802.11ac WLANs is assessed in \cite{bellalta2014channel}, where authors show significant gains under moderate neighboring activity. High-density deployments are evaluated in \cite{bellalta2016interactions}, showing the exposure to unfairness situations. Opportunistic CB under the presence of legacy users is assessed in \cite{han2016performance}. The authors in \cite{kai2017channel} propose an optimal channel allocation algorithm for DCB WLANs, showing by means of a CTMN model that the scheme with the least overlapped channels provides the highest throughput. An analytical throughput model under unsaturated traffic loads is formulated in \cite{kim2017throughput} for CB in IEEE 802.11ac and IEEE 802.11ax. Also extendable to IEEE 802.11ax WLANs, authors in \cite{khairy2018renewal} propose a model based on renewal theory for studying the performance of CB with coexisting legacy users.}
	
	\newver{While the literature on CB and general spectrum access has extensively covered a wide variety of scenarios under different assumptions, this work focuses on spatially distributed HD WLANs under non-fully-backlogged buffers. The presented results allow us to assess whether (and under what circumstances) it is convenient or not to apply overlapping approaches in front of traditional non-overlapping schemes. To the best of our knowledge, this is the first work on DCB policies considering both spatial distribution and non-saturated traffic regimes in HD WLANs.}
	
	\subsection{CB policies and CSMA/CA operation}

	\begin{figure}[t]
		\centering
		\includegraphics[width=0.98\textwidth]{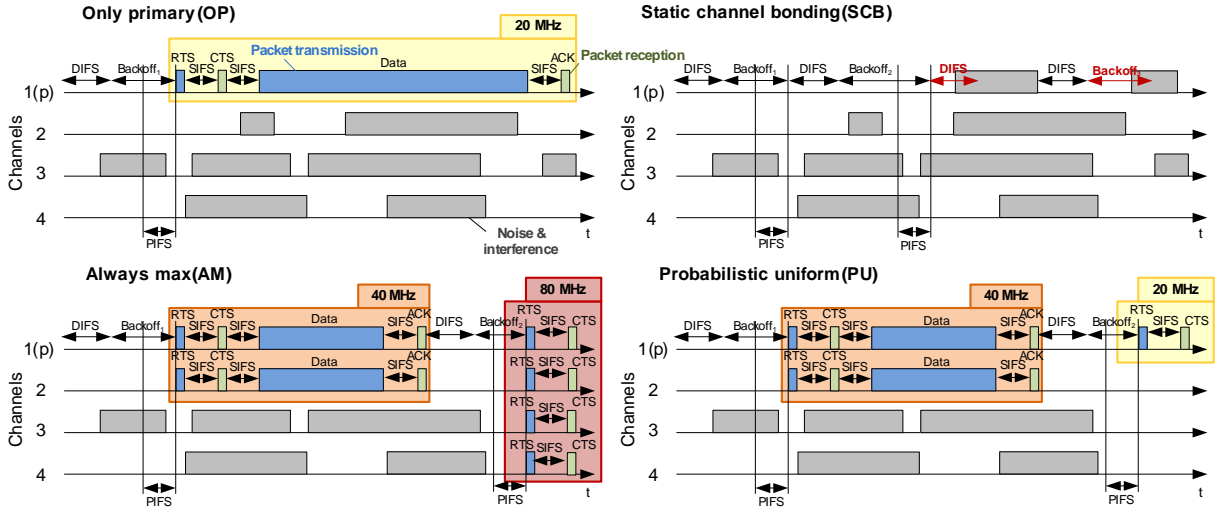}
		\caption{CSMA/CA temporal evolution of a node operating under different CB policies in a IEEE 802.11ax channelization scheme (from \cite{barrachina2018performance}). The DIFS and backoff durations in red represent that the sensed interference in the primary channel forces reseting the backoff procedure. While the duration of the legacy packets (RTS, CTS and ACK/BACK)  is the same no matter the bandwidth, the data duration is clearly reduced when transmitted in 40 \newver{and 80 MHz}.}
		\label{fig:dcb_csma_policies}
	\end{figure}

	All the aforementioned 802.11 WLAN standards operate essentially the same way when the well-known CSMA with collision avoidance (CSMA/CA) protocol is enabled. CSMA/CA works as follows: when a node $n$ belonging to a WLAN $w$ has a packet ready for transmission, it measures the power sensed in its primary channel $p_w$, and determines \newver{whether} it is idle or occupied according to the CCA level. Once $p_w$ has been detected idle \newver{during a DCF interframe space (DIFS)}, $n$ starts the backoff procedure by selecting a random initial value $b \in [0,\text{CW}-1]$, where CW is the contention window.
	After computing $b$, the node starts decreasing its counter while sensing the primary channel. Whenever the power sensed by $n$ at $p_w$ is higher than its CCA, the backoff is paused until $p_w$ is detected free again, at which point the countdown is resumed. When the backoff timer expires, the node selects the transmission channel $C_n^{\text{tx}}$ based on the set of idle basic channels\footnote{Note that, in order to include secondary channels for transmitting, a WLAN must listen \newver{to} them free during at least a Point coordination function (PCF) Interframe Space (PIFS) period before the backoff counter terminates, as shown in Figure \ref{fig:dcb_csma_policies}.} and on the implemented spectrum management rules.
	In this paper we refer to such rules as CB policies. Namely, when the backoff terminates, the node operates according to the implemented policy \newver{$\mathcal{D}$} as follows:
	\begin{itemize}
		\item \newver{\textbf{Only-primary (OP)}: if CB is not considered, we simply refer to the traditional single-channel (or only-primary operation), i.e., a node can only pick its primary channel for transmitting.}
		\item \textbf{Static channel bonding (SCB)}: exclusively picks the whole allocated channel if found entirely free (i.e., all the basic channels inside the allocated channels are free).
		\item \textbf{DCB - Always-max (AM)}: picks the widest possible combination of basic channels found free.
		\item \textbf{DCB - Probabilistic uniform (PU)}: picks with same probability any of the possible combinations of basic channels found free.
	\end{itemize}
	 Note that the computational complexity of the presented policies is very low and can be easily implemented in \newver{off-the-shelf} STAs. In fact, the most complex one is PU, which does only require to compute the outcome of a uniform random variable to determine the number of 20 MHz-channels to bond given 4 possible outcomes at the most (i.e., 1, 2, 4 or 8).
	
	The selected transmission channel is then used throughout the packet exchanges involved in a data packet transmission (i.e., RTS, CTS, data, and ACK). The duration of a successful transmission is then given by
	\begin{equation} \label{eq:t_suc}
	T_\text{suc} = T_\text{RTS}\, \Plus \, 3 \, T_\text{SIFS}\, \Plus \, T_\text{CTS} \, \Plus \, T_\text{DATA} \, \Plus \, T_\text{BACK} \, \Plus \, T_\text{DIFS} \, \Plus \, T_\text{e} \text{,}
	\end{equation} 
	where $T_\text{SIFS}$ and $T_\text{DIFS}$ are the Short Interframe Space (SIFS) and DIFS duration, respectively, and $T_\text{e}$ is the duration of an empty backoff slot. $T_\text{RTS}$, $T_\text{CTS}$, $T_\text{DATA}$ and $T_\text{BACK}$ are the transmission duration of the RTS, CTS, data, and block acknowledgment (BACK) packets, respectively. Likewise, any other node that receives an RTS in its primary channel with enough power to be decoded will enter in network allocation vector (NAV) state, which is used for deferring channel access and avoiding packet collisions (especially those caused by hidden node situations).
	
	In Figure \ref{fig:dcb_csma_policies}, the temporal evolution of a node operating under the different CB policies is shown. In this example, the node is allowed to transmit in the set of basic channels $C_w=\{1(p), 2, 3, 4\}$, where $p_w=1$ is the primary channel. While OP picks just the primary channel, the rest of policies try to bond channels in different ways.
	In this regard, SCB is highly inefficient in scenarios with partial interference. In fact, no packets can be transmitted with SCB in this example since the basic channel $\{3\} \in C_w$ is busy when both backoffs terminate. \newver{Instead}, more flexible approaches like AM and PU are able to transmit more than one frame in the same period of time.
	On the one hand, AM adapts in an aggressive way to the channel state. In this example, it is able to transmit in 40 and 80 MHz channels at the end of the first and second backoff, respectively. On the other hand, the stochastic nature of PU makes it more conservative than AM. In the example, the node could transmit in 1 or 2 basic channels with \newver{the} same probability (1/2) when the first backoff terminates. Likewise, after the termination of the second backoff, a channel composed of 1, 2 or 4 basic channels could be selected with equal probability too (1/3).
	
	
	\section{Understanding the interactions between spatially distributed WLANs} \label{sec:use_cases}
	
	In this Section\newver{,} we first analytically model the interactions given in spatially distributed WLANs under different traffic loads. Essentially, we show that the probabilities of transiting from one state to another in the generated CTMNs are determined by the CB policies of the WLANs in the network. Later, we present two toy scenarios and simulate them \newver{by means of CTMNs and \texttt{11axHDWLANsSim},\footnote{All of the source code of Komondor is open, encouraging sharing of algorithms between contributors and providing the ability for people to improve on the work of others under the GNU General Public License v3.0. The repository can be found at \url{https://github.com/wn-upf/Komondor}.} release v1.2.1b of the Komondor wireless simulator \cite{barrachina2018komondor}.}
	
	\subsection{The CTMN model for WLANs}
	
	CTMNs have been widely used \newver{in the literature} to model the behavior of WLAN networks.
	An approach which accurately models the behavior of unsaturated CSMA/CA networks operating in single-channel was introduced in \cite{laufer2013capacity}. Such \newver{a} model is extended in \cite{bellalta2016interactions} to capture the coupled dynamics of a group of overlapping WLANs using CB. Later, authors in \cite{barrachina2018performance} introduced a framework (SFCTMN) which extended the CTMN algorithm presented in \cite{faridi2016analysis} for characterizing CB policies in spatially distributed scenarios where all WLANs are saturated.
	However, to the best of our knowledge, spatial distribution effects like WLAN starvation are not considered in works studying DCB under unsaturated regimes. 
	
	Below we model such scenarios through CTMNs too. To do so, we extend the model presented in \cite{barrachina2018performance} by considering unsaturated traffic loads as proposed in \cite{laufer2013capacity}. For simplicity, we consider only downlink traffic and that each WLAN is composed by one access point (AP) and one station (STA). Hence, we simply refer to the WLAN activity as a single entity.
	
	\subsubsection{Assumptions and implications}
	
	Modeling WLAN scenarios with CTMNs requires the backoff and transmission times to be exponentially distributed. We also assume that the propagation delay between any pair of nodes is negligible. This has a main implication: the probability of \newver{slotted backoff} collisions between two or more nodes within carrier sense range is zero. Nonetheless, packet collisions resulting from the cumulated interference of simultaneous transmissions of nodes outside the carrier sense range are possible. Besides, an infinite maximum number of retransmissions per packet is assumed. Note that effect of assuming \newver{an} infinite maximum number of retransmissions is almost negligible in most of the cases because of the small probability of retransmitting a data packet more than a few times \cite{chatzimisios2004performance}.
	
	\subsubsection{States in the CTMN}
	A state $s$ in the CTMN is defined by the set of active WLANs (i.e., that are transmitting) and the basic channels selected for the transmission. The set of feasible states is represented by $\mathcal{S}$. Essentially, with slight abuse of notation, we say that a WLAN $w$ is active in state $s$, i.e., $w \in s$ if it is transmitting, and inactive otherwise. States are represented by the most left and most right basic channels used in the transmission channels of each of the active WLANs. For instance, in state $s=\text{A}_2^2 \text{B}_1^4$, there are two active WLANs: A and B. While A is transmitting in the basic channel $C_\text{A}^\text{tx} = \{2\}$ (20 MHz), B is doing so in a bonded channel $C_\text{B}^\text{tx} = \{1,2,3,4\}$ (80 MHz). The state in which there is no active WLAN is represented by $\emptyset$.
	
	A transition between two states $s$ and $s'$ in the CTMN has a corresponding transition rate $Q_{s,s'}$. For \textit{forward} transitions, the average packet transmission attempt rate is $\rho_w\lambda_w$, where $\lambda = 1/(\text{E}[B] \cdot T_\text{slot})$, being $\text{E}[B]$ the expected backoff duration in time slots. Parameter $\rho_w$ is the long-run stationary probability that WLAN $w$ has packets ready for transmission when the primary channel is sensed idle and so the backoff counter is active. Consequently, $\rho_w$ depends on the traffic load $\ell_w$ of WLAN $w$. Note that a WLAN becomes saturated ($\rho_w = 1$) whenever it is not able to carry its traffic load, i.e., whenever it generates more packets than the ones it transmits. For \textit{backward} transitions, the departure rate ($\mu$) depends on the duration of a successful transmission ($T_\text{suc}$), which in turn depends on both the data rate ($r$) given by the selected MCS and transmission channel width, and on the average data packet length ($\text{E}[L]$). Thus, we simply say that the data rate of a WLAN $w$ depends on the state of the system, which contains such information, i.e., $\mu_w(s)$.
	
	\subsubsection{Analytical performance metrics}
	The equilibrium distribution vector $\vec{\pi}$ represents the fraction of time the system spends in each state. We define $\vec{\pi}_s$ as the probability of finding the system at state $s$. Hereof, in continuous-time Markov processes with stationary distribution, $\vec{\pi}$ is given by solving the system of equations $Q \vec{\pi} = 0$, where the matrix item $Q_{s,s'}$ is the transition rate from state $s$ to $s'$.
	Once $\vec{\pi}$ is computed, estimating the average throughput experienced by each WLAN is straightforward. Specifically, the average throughput of WLAN $w$ is
	\begin{equation} \label{eq:throughput_ctmn}
	\Gamma_{w} := \text{E}[L]  \bigg(\sum_{s \in \mathcal{S}}^{}\{\gamma_{w}(s) > \text{CE} : 0, 1\}	\mu_w(s)\pi_s \big(1-\eta \big) \bigg) \text{,}
	\end{equation}
	where $\text{E}[L]$ is the expected data packet length, $\gamma_{w}(s)$ is the SINR perceived by the receiving STA in WLAN $w$ in state $s$, CE is the capture effect threshold, and $\eta$ is the MCS packet error probability.\footnote{A maximum decoding packet error rate of 10\% is usually tried to be guaranteed when selecting the MCS index in 802.11 devices.}
		
	Note that the unknown $\rho$ parameters must be obtained by solving a non-linear system of equations, which in general does not have a closed-form. As done in \cite{bellalta2016interactions}, in this work we use an iterative fixed-point approach for updating the $\rho$ values until the throughput of all the WLANs converges to their corresponding traffic load, or they become saturated.
	
	
	
	\subsection{Constructing CTMNs for CSMA/CA WLANs}

\begin{figure}[h]
		\centering
		\begin{subfigure}[t]{0.4\textwidth}
			\hfil
			\raisebox{0.32\textwidth}{%
				\includegraphics[width=0.9\textwidth,height=0.35\textwidth]{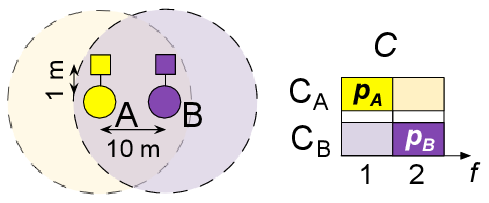}%
			}
			\caption[]{Scheme of \textit{toy scenario I}.}    
			\label{fig:scenario_I_scheme}
			\hfill
		\end{subfigure}
		~
		\begin{subfigure}[t]{0.5\textwidth}
		\begin{tikzpicture}[<->,>=stealth',shorten >=0.8pt,auto,node distance=1.8cm,
			semithick]
			\tikzstyle{every state}=[fill=white,draw=black,thick,text=black,scale=1]
			
			\node[state, label=below left:$s_1$] (S1) {$\emptyset$};
			
			\node[state, label=below:$s_3$] (S3) [above right of = S1, xshift=2.8cm, yshift=-0.5cm] {$A_{1}^{2}$};
			
			\node[state, label=below:$s_2$] (S2) [above of = S3, yshift=-0.25cm] {$A_{1}^{1}$};
			
			\node[state, label=below:$s_4$] (S4) [below right of = S1, xshift=2.8cm, yshift=0.5cm] {$B_{2}^{2}$};
			
			\node[state, label=below:$s_5$] (S5) [below of = S4, yshift=0.25cm] {$B_{1}^{2}$};
			
			\node[state, label=below:$s_6$] (S6) [below right of = S3, xshift=1.8cm, yshift=0.5cm] {$A_{1}^{1} B_2^2$};
			
			\path
			
			(S1) edge[dashed, bend left]     node[sloped, anchor=center, above]{\scriptsize $\boldsymbol{\alpha_{\text{A},\emptyset}(s_2)}\rho_\text{A}\lambda_\text{A},\mu_\text{A}(s_2)$} (S2)
			
			(S1) edge[dashed]    node[sloped, anchor=center, above]{\scriptsize \textcolor{black}{$\boldsymbol{\alpha_{\text{A},\emptyset}(s_3)}\rho_\text{A}\lambda_\text{A},\mu_\text{A}(s_3)$}} (S3)
			
			(S1) edge[dashed]   node[sloped, anchor=center, below]{\scriptsize \textcolor{black}{$\boldsymbol{\alpha_{\text{B},\emptyset}(s_4)}\rho_\text{B}\lambda_\text{B},\mu_\text{B}(s_4)$}} (S4)
			
			(S1) edge[dashed, bend right]     node[sloped, anchor=center, below]{\scriptsize$\boldsymbol{\alpha_{\text{B},\emptyset}(s_5)}\rho_\text{B}\lambda_\text{B},\mu_\text{B}(s_5)$} (S5)
			
			(S2) edge[dashed, bend left]     node[sloped, anchor=center, above]{\scriptsize $\rho_\text{B}\lambda_\text{B}, \mu_\text{B}(1)$} (S6)
			
			(S4) edge[dashed]     node[sloped, anchor=center, below]{\scriptsize $\rho_\text{A}\lambda_\text{A}, \mu_\text{A}(1)$} (S6)
			
			;
			\end{tikzpicture}
			\caption[]{CTMN corresponding to \textit{toy scenario I}.}
			\label{fig:scenario_I_ctmn}
		\end{subfigure}
		\caption[ ] {\textit{Toy scenario I.} a) WLANs A and B are inside the carrier sense range of each other with potentially overlapping basic channels 1 and 2. b) Note that certain states and transition edges are not given in the CTMN depending on the applied combination of CB policies. For instance, state $s_6$ is only reachable for the OP and PU policies, while the transitions from $s_1$ to $s_3$ and $s_5$ are not possible when OP is implemented.}
		\label{fig:mean and std of nets}
	\end{figure}

	Let us consider the toy \textit{Scenario I} shown in Figure \ref{fig:scenario_I_scheme}, which is composed of two potentially overlapping WLANs, to depict a small example of how CTMNs are constructed. The channel allocation of this scenario can be defined as $\mathcal{C}\text{: } C_\text{A} = \{1(p), 2\}$ with $p_\text{A} = 1$, and $C_\text{B} = \{1, 2(p)\}$ with $p_\text{B} = 2$. That is, there are two basic channels in the system, and the set of valid transmission channels according to the IEEE 802.11ax channel access scheme is $\{\{1\}, \{2\}, \{1,2\}\}$. We say that both WLANs are potentially overlapping because they are inside the carrier sense range of each other and thus their signals will overlap when transmitting in the same channel at the same time $t$, i.e., when $C^{\text{tx}}_\text{A}(t) \cap C^{\text{tx}}_\text{B}(t) \neq
	\emptyset$. In this case, due to the primary channel allocation, A and B will only overlap when both transmit in their whole allocated channel $\{1,2\}$.
	
	Different feasible states and \textit{forward} transitions may exist in the CTMN depending on the implemented CB policies. Every feasible transition is weighted by a transition probability vector $\alpha_{w,s}(s')$ whose elements determine the probability of WLAN $w$ to transit from state $s$ to $s'$.	
	Table \ref{table:cb_policy_effect_alphas} collects the number of feasible states ($|\mathcal{S}|$) and transition probabilities that are given for each of the studied CB policies in \textit{Scenario I}. The corresponding CTMNs are shown in Figure \ref{fig:scenario_I_ctmn}.

	\begin{table}[h]
		\centering
		\footnotesize
		\caption{Transition probabilities from state $\emptyset$ of WLANs A and B in \textit{toy scenario I} for different CB policies.}
		\label{table:cb_policy_effect_alphas}
		\begin{tabularx}{.6\textwidth}{CCCCCC}
			\toprule
			
			$\boldsymbol{\mathcal{D}}$ & 
			$\boldsymbol{|\mathcal{S}|}$ & 
			$\boldsymbol{\vec{\alpha}_{\text{A},\emptyset}(s_2)}$ &
			$\boldsymbol{\vec{\alpha}_{\text{A},\emptyset}(s_3)}$ & $\boldsymbol{\vec{\alpha}_{\text{B},\emptyset}(s_4)}$ & $\boldsymbol{\vec{\alpha}_{\text{B},\emptyset}(s_5)}$ \\
			
			\midrule
			OP            & 4               & 1.0             & 0.0             & 1.0             & 0.0             \\ 
			SCB           & 3               & 0.0             & 1.0             & 0.0             & 1.0             \\ 
			AM            & 3               & 0.0             & 1.0             & 0.0             & 1.0             \\ 
			PU            & 6               & 0.5             & 0.5             & 0.5             & 0.5             \\ 
			\bottomrule
		\end{tabularx}
	\end{table}

	For instance, with OP, since WLANs are only allowed to transmit in their primary channel, the CTMN can only transit from state $\emptyset$ to states $\text{A}_1^1$ or $\text{B}_2^2$, i.e., $\alpha_{\text{A},\emptyset}(s_2)=\alpha_{\text{B},\emptyset}(s_4) = 1$. Instead, with SCB, WLANs can only transmit in their complete allocated channel, thus, when being in state $\emptyset$ the CTMN transits to the \textit{all or nothing} states $\text{A}_1^2$ or $\text{B}_1^2$, i.e., $\alpha_{\text{A},\emptyset}(s_3)=\alpha_{\text{B},\emptyset}(s_5) = 1$. Notice that in this particular case AM generates the same transition probabilities (and respective average throughput) than SCB because whenever the WLANs have the possibility to transmit -- which only happens when the CTMN is in state $\emptyset$ -- they pick the widest channel available, i.e., $\{1,2\}$. Finally, PU picks uniformly at random any of the possible transitions when the backoff terminates in $\emptyset$, i.e., $\alpha_{\text{A},\emptyset}(s_2)=\alpha_{\text{A},\emptyset}(s_3) = 1/2$ and $\alpha_{\text{B},\emptyset}(s_4)=\alpha_{\text{B},\emptyset}(s_5) = 1/2$, respectively.
	
	\subsection{Empirical performance metrics and toy evaluation setup}
	
	Note that, even though the analytical expression of the throughput by CTMN (\ref{eq:throughput_ctmn}) is pretty accurate \cite{barrachina2018performance}, there are other performance metrics hard to capture with enough accuracy because of the required assumptions like the nonexistence of backoff collisions. That is why in this work we rely on the event-based wireless network simulator \texttt{11axHDWLANsSim} \cite{barrachina2018komondor}. The performance metrics considered in this work are \newver{defined} as follows:
	
	\begin{itemize}
		\item \textbf{Throughput $\boldsymbol{\Gamma}$}: total number of data bits successfully sent (i.e., acknowledged) during the observation time. That is, only the useful data (i.e., no headers) of each of the transmitted frames is considered for computing the throughput.
		\item \textbf{Access delay $\boldsymbol{\delta}$}: average duration between two consecutive channel accesses whenever there is backlogged data.
		\item \textbf{Packet delay $\boldsymbol{d}$}: average delay between a packet arrival (insertion in the buffer queue) and its corresponding acknowledgment after being transmitted.
		\item \textbf{Drop ratio $\boldsymbol{\varphi}$}: ratio of packets that are dropped by the buffer. A packet is dropped if it is generated when the queue of the buffer is full (i.e., when the buffer already has $N_b$ packets at the queue).
		\item \textbf{No. of aggregated packets per frame $\boldsymbol{n_a}$}: average number of aggregated packets per frame. A frame can contain up to $N_a$ packets.
	\end{itemize}
	
	The parameters of the simulation setups evaluated in this work\footnote{For the sake of saving space, the full details of the evaluation setups (e.g., nodes positions) and corresponding results of the scenarios considered through the paper are detailed in \url{https://github.com/sergiobarra/data_repos/tree/master/barrachina2018tooverlap}.} are collected in Table \ref{table:appendix_table} of the Appendix, which correspond to the IEEE IEEE 802.11ax simulation setup presented in \cite{barrachina2018performance}. However, for the sake of simplicity, in \newver{these} toy scenarios we consider for the moment no MCS error rate ($\eta = 0$) and highest MCS corresponding to 1024-QAM 5/6. Regarding the traffic load, note that we consider that a WLAN $w$ generates a data packet every  $t_w \sim \text{Exponential}(1 / \ell_w)$, following a Poisson process.
	
	\subsection{Toy scenario I: to overlap or not \newver{to overlap}?}
	
\begin{figure}[t]
		\centering
		\includegraphics[width=0.5\textwidth]{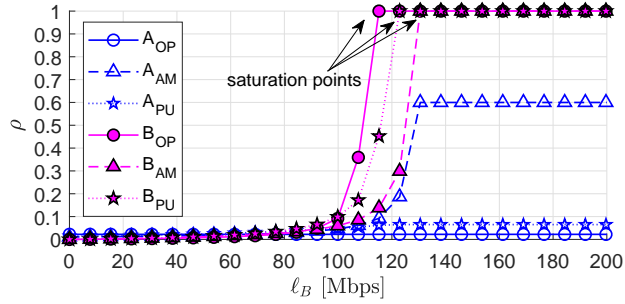}
		\caption{Saturation in \textit{toy scenario I}.  The long-run stationary probability $\rho$ is estimated through SFCTMN. Traffic load of WLAN A is fixed to $\ell_A = 76.8$ Mbps. WLAN B gets saturated at different traffic loads $\ell_B$ depending on the CB policy.}
		\label{fig:toy_scenario_I_rho}
	\end{figure}

\begin{figure}[t]
		\centering
		\includegraphics[width=1\textwidth]{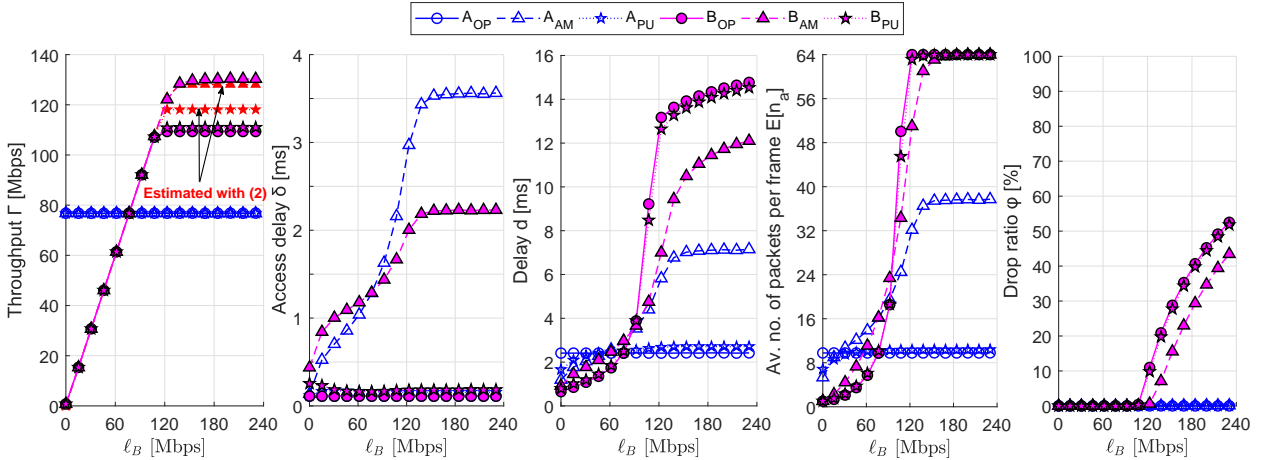}
		\caption{Performance metrics of \textit{toy scenario I}. Results correspond to \texttt{11axHDWLANsSim} simulations of 1000 seconds each. Traffic load of WLAN A is fixed to $\ell_A = 76.8$ Mbps. The throughput curves in red correspond to the only values obtained by the CTMN model (\ref{eq:throughput_ctmn}) that do not completely match the simulator results.}
		\label{fig:toy_scenario_I_plot}
	\end{figure}

	In Figure \ref{fig:toy_scenario_I_rho} there is plotted the long-run stationary probability $\rho$ of both WLANs when operating under different policies and traffic loads. Likewise, in Figure \ref{fig:toy_scenario_I_plot}, we plot the average throughput, access delay, packet delay, drop ratio and number of aggregated data packets per frame. While we keep the traffic load of A constant to $\ell_\text{A} = 76.8$ Mbps, the load of B is the x-axis independent variable $\ell_\text{B} \in [0, 240]$ Mbps. We assume that both WLANs implement exactly the same policy in each case.

	Given the duration of a successful slot in a CSMA/CA IEEE 802.11ax network (\ref{eq:t_suc}), the maximum capacity for a successful transmission of a frame containing $N_a$ packets, i.e., $r=N_a L_D/T_\text{suc}(N_a)$, using MCS 11 is $r_{20} = 109.71$ Mbps for single-channel (20 MHz) transmissions, and $r_{40} = 207.18$ Mbps for two bonded channels (40 MHz) transmissions. Thus, even in isolation, whenever the traffic load of a WLAN surpasses that rates, it gets saturated. Note also the effect of the overhead introduced by the PHY and MAC layers since the raw transmission data rates provided by 1024-QAM 5/6 are $r_{20}^* = 121.9$ and $r_{40}^* = 243.8$ Mbps, for 20 and 40 MHz, respectively. 
	
	The saturation points of WLAN B are shown in Figure \ref{fig:toy_scenario_I_rho}, where for OP it gets saturated at approximately $\ell_B \approx r_{20}$. Instead, regarding A's saturation point, we note that, as single-channel capacity already copes with $\ell_\text{A}$ (i.e., $\ell_\text{A} < r_{20}$), it never gets saturated (i.e., $\rho_\text{A} < 1$) no matter neither the policy selected nor $\ell_\text{B}$.\footnote{Note that $\ell < r$ is a mandatory condition in order to ensure unsaturated regimes. The reason lies in the overheads caused by the headers, control packets and inter frame spaces of the MAC layer.} As expected, with AM, B gets saturated for higher $\ell_\text{B}$ since more frames can be transmitted per unit of time. Note that in isolation, B would saturate for a $\ell_\text{B}$ close to $r_{40}$. In this case, however, the whole channel is shared with A when both implement AM and saturation is reached at a lower value $\ell_\text{B} \approx 130$ Mbps.

	
	In terms of throughput, the higher the traffic load required to saturate a WLAN, the higher its potential value. That is, AM provides the highest $\Gamma_\text{B}$ for high $\ell_\text{B}$, while any policy combination copes with $\ell_\text{A}$ (i.e., $\ell_\text{A} = \Gamma_\text{A}$).
	Regarding the CTMN model, all the throughput estimations completely match the simulator results with exception of the slight difference given in the $\text{B}_\text{AM}$ and $\text{B}_\text{PU}$ curves. On the one hand, the main reason lies in the fact that the CTMN model assumes that all the frames contain exactly $N_a$ packets, while the simulator has not such a restriction. Thus, frames containing less than $N_a$ packets are completely possible in the simulations conducted. This effect is specially noticeable at curve $\text{B}_\text{PU}$. On the other hand, while simultaneous slotted backoff terminations are not captured by the CTMN model, the simulator does so. Hence, since in this particular scenario concurrent transmissions are decodable due to the proximity AP-STA, $\text{B}_\text{AM}$ is slightly smaller than the simulated one.
	
	As it occurs with the throughput, for high $\ell_B$, it is more convenient for B to use the aggressive CB approach provided by AM in order to decrease the delay. However, interestingly, we note that for low $\ell_B$, OP is the best policy since the delay to access the channels is significantly reduced compared to AM. The reason is that with AM the two WLANs must share the channel as they transmit using the full 40 MHz spectrum. Consequently, backoff counters get frozen during larger periods of time and the delay between consecutive channel accesses increases accordingly. This effect can also be seen in the average number of packets aggregated per frame, where, for low traffic loads, AM aggregates more packets on average since the buffer of one WLAN is able to be filled with more packets during the transmission of the other. In turn, when the backoff expires, larger frames are sent.
	
	In this particular scenario, we see that an overlapping approach is the best both in terms of delay and throughput when the traffic load is moderate to high. Instead, for low traffic loads, \newver{the} delay is reduced with OP, since it avoids overlaps making channel access independent on the other WLAN's activity.
	
	\subsection{Toy Scenario II: drawbacks of overlapping} \label{subsec:toy_scenario_2}
	
\begin{figure}[t]
		\centering
		\includegraphics[width=0.55\linewidth]{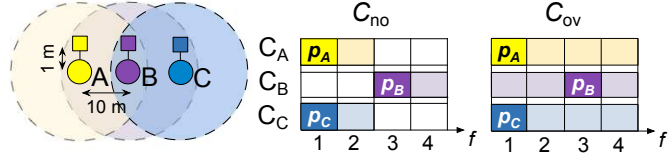}
		\caption{\textit{Toy scenario II}. A \textit{neighbor overlapping} network where two channel allocations are considered.}
		\label{fig:scenario_II_scheme}
	\end{figure}

	\textit{Toy scenario II} shown in Figure \ref{fig:scenario_II_scheme} comprises a network of three WLANs where the central one (B) is in the carrier sense range of the other two (A and C). Instead, A and C are outside the carrier sense of each other (i.e., the edge WLANs never overlap in any basic channel). All the WLANs implement AM. We consider two different channel allocations for comparing the non-overlapping vs. overlapping approaches, respectively:
	\begin{itemize}
		\item $\boldsymbol{\mathcal{C}_\text{no}}$: $C_\text{A} = C_\text{C} = \{1(p),2\}$ and $C_\text{B} = \{3(p),4\}$.
		\item $\boldsymbol{\mathcal{C}_\text{ov}}$: $C_\text{A} = C_\text{C} = \{1(p),2,3,4\}$ and $C_\text{B} = \{1,2,3(p),4\}$.
	\end{itemize}

	\begin{figure}[t]
		\centering
		\begin{subfigure}[b]{0.5\textwidth}
			
			\begin{tikzpicture}[<->,>=stealth',shorten >=1pt,auto,node distance=1.95cm,
			semithick]
			\tikzstyle{every state}=[fill=white,draw=black,thick,text=black,scale=0.8]
			
			\node[state, label=below:$s_1$]    (S1)                    {$\emptyset$};
			
			\node[state, label=below:$s_3$]    (S3)[right of = S1, xshift=0.5cm]		{$\text{B}_3^4$};
			
			\node[state, label=below:$s_2$]    (S2)[above of = S3]		{$\text{A}_1^2$};
			
			\node[state, label=below:$s_4$]    (S4)[below of = S3]		{$\text{C}_1^2$};
			
			\node[state, label=below:$s_6$]    (S6)[right of = S3, xshift=0.5cm]		{$\text{A}_1^2\text{C}_1^2$};
			
			\node[state, label=below:$s_5$]    (S5)[above of = S6]		{$\text{A}_1^2\text{B}_3^4$};
			
			\node[state, label=below:$s_7$]    (S7)[below of = S6]		{$\text{B}_3^4\text{C}_1^2$};
			
			\node[state, label=below:$s_8$]    (S8)[right of = S6, xshift=0.5cm]		{$\text{A}_1^2\text{B}_3^4\text{C}_1^2$};
			
			\node[]    (S99)[above of = S1,]		{$\boldsymbol{\mathcal{C}_\text{no}}$};
			
			\path
			
			(S1) edge[bend left] (S2)
			
			(S1) edge (S3)
			
			(S1) edge[bend right] (S4)
			
			(S2) edge  (S5)
			
			(S2) edge (S6)
			
			(S3) edge (S5)
			
			(S3) edge (S7)
			
			(S4) edge (S6)
			
			(S4) edge  (S7)
			
			(S5) edge  (S8)
			
			(S6) edge    (S8)
			
			(S7) edge     (S8);
			
			\end{tikzpicture}
			\caption{Non-overlapping channel allocation $\mathcal{C}_\text{no}$.}
			\label{fig:scenario_II_ctmn_no}
		\end{subfigure}
		\begin{subfigure}[b]{0.4\textwidth}
			\begin{tikzpicture}[<->,>=stealth',shorten >=1pt,auto,node distance=1.95cm,
			semithick]
			\tikzstyle{every state}=[fill=white,draw=black,thick,text=black,scale=0.8]
			
			\node[state, label=below:$s_1$]    (S1)                    {$\emptyset$};
			
			\node[state, label=below:$s_3$]    (S3)[right of = S1, xshift=0.5cm]		{$\text{B}_1^4$};
			
			\node[state, label=below:$s_2$]    (S2)[above of = S3]		{$\text{A}_1^4$};
			
			\node[state, label=below:$s_4$]    (S4)[below of = S3]		{$\text{C}_1^4$};
			
			\node[state, label=below:$s_5$]    (S5)[right of = S3, xshift=0.5cm]		{$\text{A}_1^4\text{C}_1^4$};
			
			\node[]    (S99)[above of = S1,]		{$\boldsymbol{\mathcal{C}_\text{ov}}$};
			
			\path
			
			(S1) edge[bend left] (S2)
			
			(S1) edge (S3)
			
			(S1) edge[bend right] (S4)
			
			(S2) edge[bend left]  (S5)
			
			(S4) edge[bend right] (S5);

			\end{tikzpicture}
			\caption{Overlapping channel allocation $\mathcal{C}_\text{ov}$.}
			\label{fig:scenario_II_ctmn_ov}
		\end{subfigure}
		\caption{CTMNs of \textit{Toy scenario II}.}
	\end{figure}
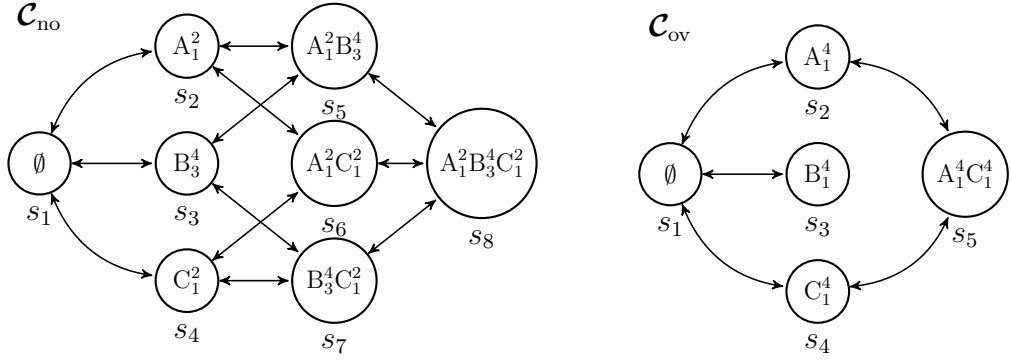

\begin{figure}[t]
		\centering
		\includegraphics[width=0.72\textwidth]{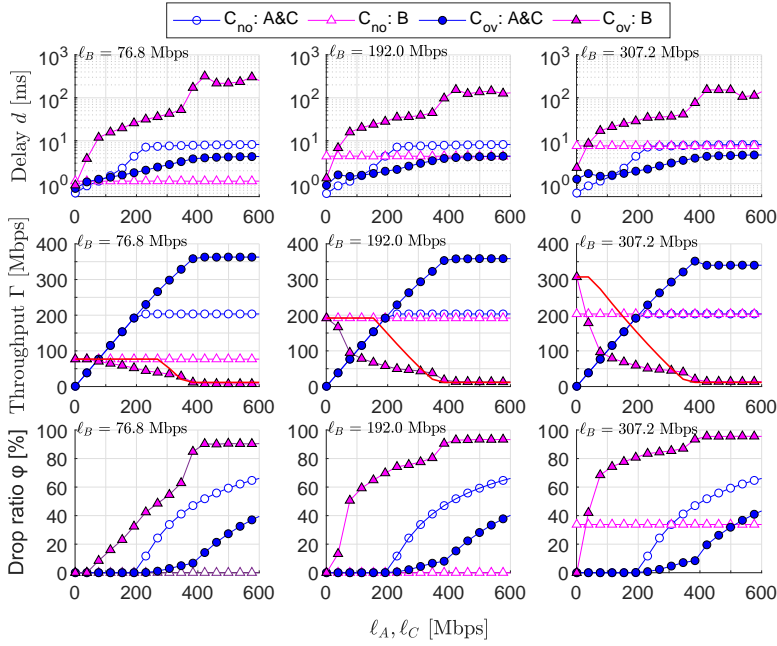}
		\caption{Performance metrics of \textit{toy scenario II}. Results correspond to simulations of 1000 seconds each. \newver{The throughput curves in red correspond to the unique case (\textit{C$_\text{ov}$:B}) where the CTMN model does not match the simulator.}}
		\label{fig:toy_scenario_II_plot}
	\end{figure}

	Note that, as shown in Figure \ref{fig:scenario_II_ctmn_no} and Figure \ref{fig:scenario_II_ctmn_ov}, different states are reached depending on the channel allocation of the WLANs. On the one hand, $\mathcal{C}_\text{no}$ allows any combination of concurrent transmissions by sacrificing potential allocated bandwidth. On the other hand, WLANs must \newver{contend} for the channel when $\mathcal{C}_\text{ov}$ is allocated. In turn, their data transmission rate is approximately doubled with respect to $\mathcal{C}_\text{no}$ (i.e., $r_{80} \approx 2 \, r_{40}$).	
	In Figure \ref{fig:toy_scenario_II_plot}, the packet delay, throughput\newver{,} and drop ratio experienced by the WLANs under different traffic loads is shown. Note that A and C behave exactly the same way since they are symmetrically deployed and have same channel allocation. We evaluate the aforementioned performance metrics for three different values of $\ell_\text{B}$ (i.e., 76.8, 192.0 and 307.2 Mbps) and a several values of $\ell_A=\ell_C=\ell_e$ in the range $[0, 600]$ Mbps.
	
	As expected, for the non-overlapping channelization ($\mathcal{C}_\text{no}$), there is no dependence among WLANs. Essentially, performance is just \newver{a} consequence of the WLANs' own traffic load. Thus, saturation is reached when $\ell$ approximates $r_{40}$ since each WLAN is allocated two basic channels. This saturation effect can be seen in the throughput and drop ratio curves in Figure \ref{fig:toy_scenario_II_plot}. Note that packets start to be dropped in saturation, i.e., when $\rho \approx 1 \rightarrow \varphi > 0$. Besides, due to the fact that channels do not overlap, the maximum throughput reachable by any WLAN is slightly less than $r_{40}$.
	
	Regarding the overlapping channel allocation ($\mathcal{C}_\text{ov}$), results show that B's performance is really deteriorated when the traffic load of A and C increases. Essentially, while A and C can transmit at the same time whenever B is not active, B can only do so when neither A nor C are active. This is a clear case of unfair WLAN starvation. Namely, the larger $\ell_e$, the fewer the transmission opportunities for B, \newver{like} A and C transmit during the majority of \newver{the} time. As a consequence of such flow-in-the-middle (FIM) starvation, B suffers from high delay, low throughput\newver{,} and high drop ratio. Interestingly, $\ell_\text{B}$ does not practically affect to $d_\text{A}$ or $d_\text{C}$, since B starves even for $l_\text{B} = 76.8$ Mbps when $\ell_e$ is high. Instead, for low $\ell_e$, the overlapping setup is more convenient for B when its traffic load is high (see performance for $l_\text{B} = 307.2$ Mbps).
	
	\newver{As for the throughput estimation of the CTMN model, we note that only WLAN B at the overlapping allocation shows different numerical results, even though following the same trend as the simulator.
	The main reason lies in two aspects. First, a fixed number of $N_a$ data payloads aggregated per frame is assumed in the CTMN model for every WLAN. Instead, the simulator realistically aggregates \textit{up to} $N_a$ according to the buffer status of each WLAN. Thus, in the model, A and C access the channel less frequently than in the simulator for low-moderate loads but do so during larger periods to transmit larger frames. Instead, since A and C are not saturated in the simulator for such traffic load, shorter but more frequent frames are transmitted. Second, the simulator's implementation of the NAV state -- resulting from properly decoding an RTS or CTS of a WLAN within the carrier sense -- is not captured by the CTMN. These make WLAN B at the simulator to have fewer chances to access the channel, thus experiencing a significantly reduced (and more representative) throughput.
	While analytical models such as the presented provide important insights into the system states and behavior of WLANs, inaccuracies when capturing complex scenarios are hard to prevent. Solving those inaccuracies in the model may be not possible or require complex extensions. For instance, in our case, we should accurately represent the buffer occupancy for each node, which would add many more dimensions and states to the CTMN.}
	
	In conclusion, we see that, in terms of delay, the non-overlapping channel allocation is the most convenient for keeping a fair deployment where all WLANs are capable to cope with low to moderate traffic loads. Besides, while an overlapping approach is really convenient for the edge WLANs in terms of throughput and delay, it is not the case for the central WLAN B\newver{, which actually starves for high edge traffic load}. Nonetheless, \newver{when neighboring activity is low}, B's performance is improved for high $\ell_B$ since it is able to use the full frequency spectrum like in isolation.
	
	The presented toy scenarios \newver{suggest} that there is not a unique spectrum allocation approach that suits all the cases. In fact, WLANs' performance \newver{depends} on multiple factors like spatial distribution and traffic loads, but also on the metric objective to be optimized, which may be designed to foster individual or collaborative behaviors. Nonetheless, we have seen that, as a rule of thumb, AM and overlapping channel allocations are the most convenient for improving the \newver{average} performance. In turn, with AM we run the risk of jeopardizing WLANs that may fall in FIM starvation or \newver{hidden node} situations with higher probability, resulting in less fair scenarios.

	
	\section{Performance evaluation in HD scenarios}	\label{sec:evaluation}
	
	In this section\newver{,} we analyze the performance of DCB in two different types of IEEE 802.11ax WLAN deployments. Namely, we first assess the impact of the node density in networks with homogeneous CB policies and traffic load. We then discuss what is the optimal CB policy that a WLAN should pick in a completely random HD deployment. The IEEE 802.11ax configuration and other setup parameters used in the following simulations are detailed in Table \ref{table:appendix_table} in the Appendix A.
	
	\subsection{Node density effect on CB policies}
		
	In order to get insights \newver{into} the node density effect on the efficiency of CB, we \newver{now assess} the performance of single-channel (non-overlapping approach) and DCB (overlapping approach) in a network consisting of 6 WLANs randomly located in a square map of different sizes: 20x20, 40x40 and 80x80 m$^2$. All the WLANs are set with the same policy in each case (OP or AM). \newver{In addition}, the same traffic load is assumed for all the WLANs. The minimum distance between any two APs is set to $d_{\text{AP-AP}}^\text{min} = 8$ m and each WLAN is located uniformly at random at a distance $d_\text{AP-STA} \in [d_\text{AP-STA}^\text{min}, d_\text{AP-STA}^\text{max}] = [1, 4]$ m from the AP.
	Regarding the channel allocation, all the WLANs are set with random primary channel in the eight basic channels considered in the system (i.e., $p_\text{w} \sim U[1,8], \forall w$). The set of allocated basic channels is set to the maximum allowed $C_w = \{1,...,8\} \forall w$ for contiguous spectrum transmissions.\footnote{\newver{Note that allocating the whole bandwidth to all the WLANs is an interesting extreme case when precalculated channel allocation is overlooked.}}
	Specifically, we generate $N_\text{D} = 200$ deployments following the aforementioned conditions for each of the $N_\text{P}=2$ policies considered. Besides, we evaluate each policy for $N_\ell = 11$ values of the homogeneous traffic load ranging from 0.768 to 150 Mbps in the $N_\text{M} = 3$ maps of different sizes. The \newver{simulated} time of each of the $N_\text{D} \times N_\text{P} \times N_\ell \times N_\text{M} = 13200$ scenarios is 20 seconds.
	
\begin{figure}[t]
		\centering
		\begin{subfigure}[b]{0.475\textwidth}
			\centering
			\includegraphics[width=\textwidth]{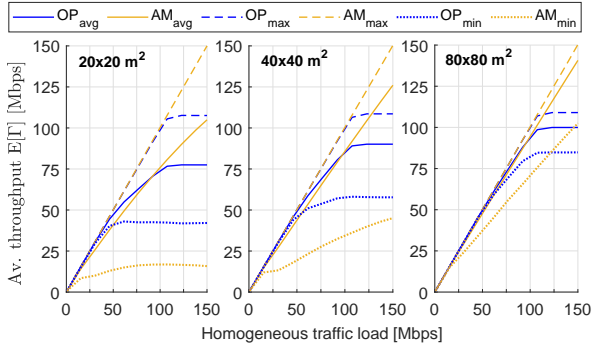}
			\caption{Average, minimum, and maximum throughput. \newver{The legend's subscripts \textit{avg}, \textit{max} and \textit{min} refer to the average, maximum, and minimum of the performance metrics.}}
			\label{fig:random_rect_throughput}
		\end{subfigure}
		\hfill
		\begin{subfigure}[b]{0.475\textwidth}  
			\centering 
			\includegraphics[width=\textwidth]{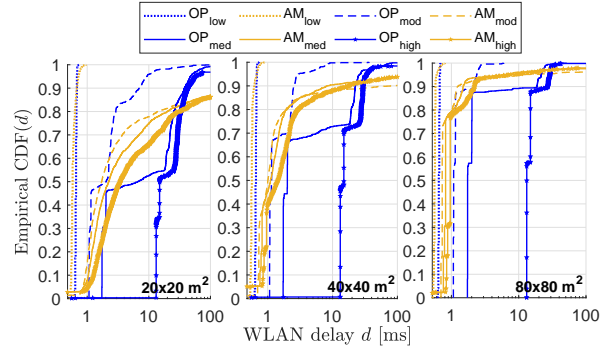}
			\caption{\newver{Empirical CDF of the individual delay. The legend's subscripts \textit{low}, \textit{mod}, \textit{med}, and \textit{high} refer to $\ell=0.768, 30.72, 61.44, 122.88$ Mbps, respectively.}}
			\label{fig:random_rect_delay}
		\end{subfigure}
		\vskip\baselineskip
		\begin{subfigure}[b]{0.475\textwidth}   
			\centering 
			\includegraphics[width=\textwidth]{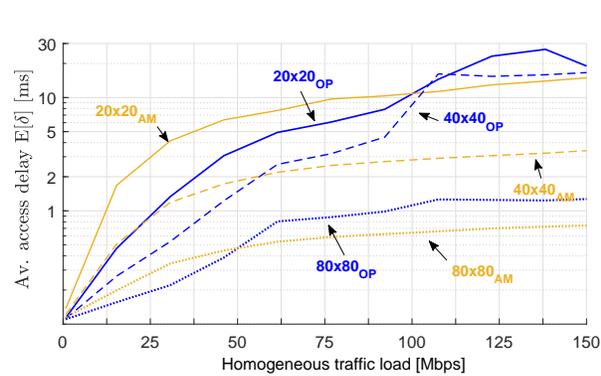}
			\caption{Average access delay for OP and AM in the three map deployments considered.}
			\label{fig:random_rect_wait_time}
		\end{subfigure}
		\quad
		\begin{subfigure}[b]{0.475\textwidth}   
			\centering 
			\includegraphics[width=\textwidth]{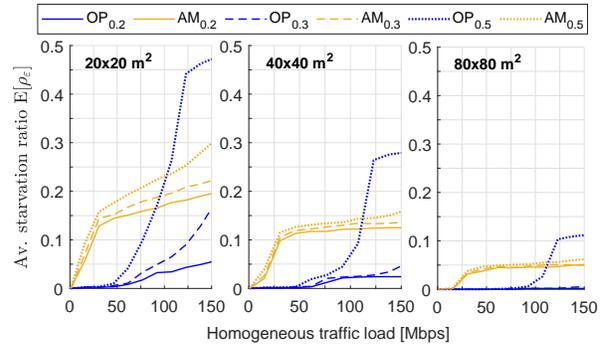}
			\caption{Average starvation ration according to three starvation thresholds. Subscripts in the legend represent the value of $\varepsilon$.}
			\label{fig:random_rect_starvation}
		\end{subfigure}
		\caption{Node density effect on network performance. 6 WLANs with same traffic load and CB policies are deployed in squared maps of different areas.}
		\label{fig:6Wlans}
	\end{figure}

	\newver{In Figure~\ref{fig:random_rect_throughput} it is shown the average, maximum and minimum throughput experienced by the network.} We note that OP has a clear limitation regarding the maximum achievable value, which leads to saturation even in the \newver{least} dense map (80x80 m$^2$). In fact, such maximum is quasi-independent of the node density and corresponds to the maximum \newver{effective} data rate provided by 20 MHz transmissions $r_{20} = 109.71$ Mbps. Instead, for mid to low dense scenarios, AM is able to cope with the traffic load in the majority of the cases as reflected by the average and maximum throughput curves of maps 40x40 and 80x80 m$^2$. \newver{As indicated by the minimum throughput, we find more situations with precarious performance in AM, which corroborates the \textit{risky} nature of aggressive DCB policies. That is, while AM is convenient on average, it is more prone to generate unfair scenarios where at least one WLAN \newver{experiences} poor performance.}

	\newver{The cumulative distribution function (CDF) of the packet delay experienced by the WLANs is plotted in Figure \ref{fig:random_rect_delay}. In contrast to the throughput, studying the CDF is convenient for the packet delay since its average value may explode even when just one of the WLAN starves. As expected, the higher the traffic load, the less the probability of achieving small delays regardless of the selected policy. As suggested in previous sections, we confirm that OP leads to a smaller delay than AM for low loads when WLANs are likely to overlap (see 20x20 m$^2$ map). Instead, for higher loads, since OP is by default constrained to the maximum data rate provided by a 20 MHz channel, it is more probable to achieve acceptable delays with AM. For less dense scenarios, the risk of generating unfair situations is importantly reduced in AM (see 80x80 m$^2$ map). We note that the packet delay matches completely with the average access delay shown in Figure \ref{fig:random_rect_wait_time}. A key aspect to consider in this regard is the effect on the average backoff duration in presence of multiple hidden node collisions, leading to a significant increase of the contention window.}
	
	\newver{To conclude our observations on} the fairness and intrinsic risk of DCB, we study the average starvation ratio. We say that a WLAN $w$ starves if it is not able to successfully transmit a certain fraction $\varepsilon$ of its traffic load $\ell_w$. Specifically, $w$ starves if its average throughput $\Gamma_{w}$ is less than the selected starvation threshold, i.e., when $\Gamma_{w} < \varepsilon \ell_w$. The starvation ratio $\rho_\varepsilon$ of a particular scenario is computed as the fraction of starving WLANs. For instance, if 2 of 6 WLANs are found to be starving, the corresponding starvation ratio would be 2/6. The average value of this ratio is plotted in Figure \ref{fig:random_rect_starvation} for the different maps.
	\newver{As expected, the higher the traffic load and/or map density, the higher the starvation ratio for both policies.}
	\newver{Looking at the low-mid traffic loads we confirm that OP outperforms AM when it comes to avoid unfair scenarios in uncontrolled deployments. Indeed, OP completely avoids starvation for each of the thresholds when the traffic load is low. This contrasts with AM, which outperforms OP for scenarios under very high loads, but is not able to properly  cope with the FIM and hidden node situations in low density deployments.}

	\subsection{Optimal individual policy in uncontrolled HD WLANs}
	
\begin{figure}[t]
		\centering
		\includegraphics[width=0.64\textwidth]{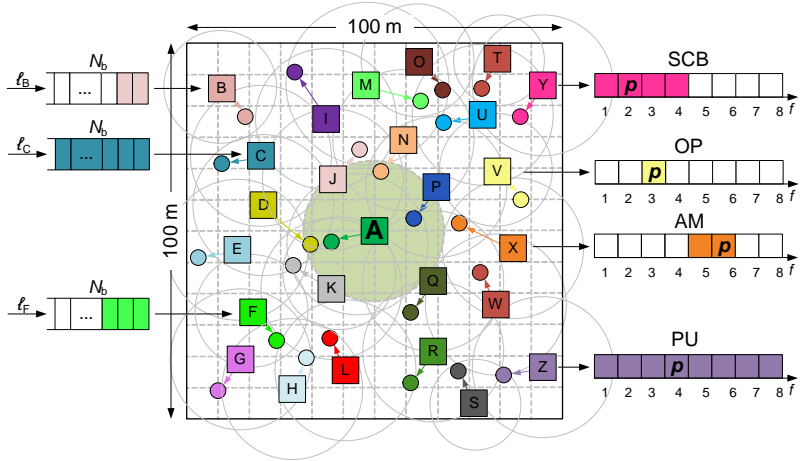}
		\caption{Network deployment with WLAN A placed in the
			middle and the rest 24 WLANs spread uniformly at
			random in a 100x100 m$2$ \newver{area}.}
		\label{fig:map_central}
	\end{figure}

	\newver{In this subsection we discuss what is the optimal CB policy that a particular WLAN (A) should locally implement for improving either its own throughput or delay, i.e., $\mathcal{D}^*_{\Gamma,\text{A}}= \argmax_\mathcal{D} \Gamma_\text{A}$ or $\mathcal{D}^*_{d,\text{A}}= \argmin_\mathcal{D} d_\text{A}$, respectively.}
	As shown in Figure \ref{fig:map_central}, we consider a 100 x 100 m$^2$ \newver{area} with WLAN A located at the center, and 24 WLANs spread uniformly at random in the area with the single condition that any pair of APs must be separated at least $d_{\text{AP-AP}}^\text{min} = 10$ m. The STA\footnote{Note that in the considered scenarios, having one or multiple STAs per AP does not significantly affect the obtained average results. \newver{The main reason is that STAs are randomly placed near the AP and the destination is selected at random for each frame transmission}. Therefore, the unique effect of considering more STAs is a probable slight decrease (increase) on the average throughput because of the performance anomaly resulting from the lower (higher) MCS picked by the STA placed farthest from (closest to) the AP.} of each WLAN is located also uniformly at random at a distance $d_\text{AP-STA} \in [d_\text{AP-STA}^\text{min}, d_\text{AP-STA}^\text{max}] = [1, 5]$ m from the AP.
	
	Regarding the channel allocation, all the WLANs are set with random primary channel in the eight basic channels considered in the system (i.e., $p_\text{w} \sim U[1,8], \forall w$). The set of allocated basic channels is assigned uniformly at random as well. That is, the number of allowed basic channels for transmitting is $|C_w| \sim U\{1,2,4,8\}, \forall w\neq\text{A}$, with the exception of WLAN A, which is allocated the widest channel (i.e., $C_\text{A} = \{1,...,8\}$). Besides, we consider now bursty traffic dependent on the average traffic load ($\ell$), where a burst of $n_b = 10$ packets is generated each $t_b \sim \text{Exponential}(n_b / \ell)$ in order to provide more realistic traffic patterns.
	
	While the CB policies of the rest of WLANs are also set uniformly at random (i.e., they implement OP, SCB, AM or PU with same probability 1/4), A is fixed to a desired policy. Specifically, we generate $N_\text{D} = 100$ deployments following the aforementioned conditions for each of the $N_\text{P}=4$ policies that A can implement. Besides, we evaluate each policy for $N_\ell = 13$ values of A's traffic load ranging from 0.768 to 184.32 Mbps (i.e., from 64 to 15360 packets/s). The rest of WLANs are set with random average traffic load inside such \newver{a} range, i.e., $\ell_w \sim U[0.768,184.32], \forall w\neq\text{A}$. Hence, we simulate $N_\text{D} \times N_\text{P} \times N_\ell = 5200$ scenarios. The simulation time of each scenario is 10 seconds.
	
\begin{figure}[t]
		\centering
		\begin{subfigure}[b]{0.475\textwidth}
			\centering 
			\includegraphics[width=\textwidth]{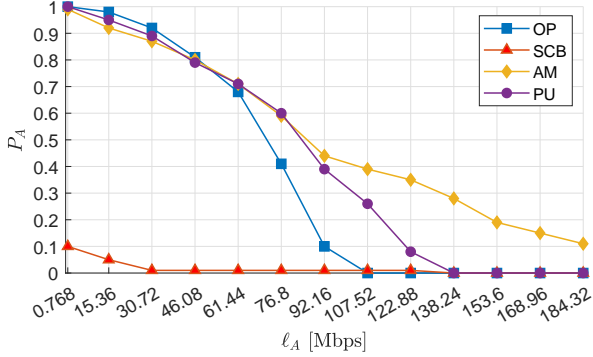}
			\caption[]{Probability that A is able to successfully transmit all its traffic load. A margin error $\epsilon_\Gamma = 0.05$ is considered.}
			\label{fig:probability_throughput_load_similar}
		\end{subfigure}
		~
		\begin{subfigure}[b]{0.475\textwidth}  
			\centering 
			\includegraphics[width=\textwidth]{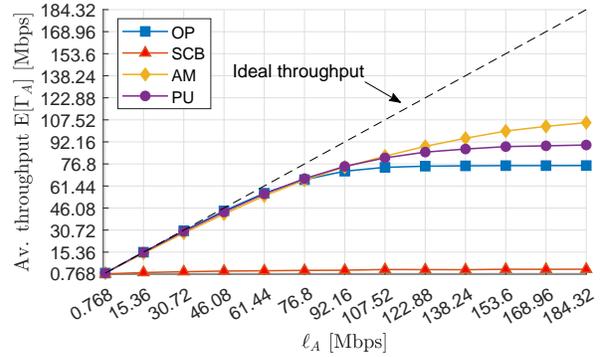}
			\caption[]{Traffic load effect on the average throughput experienced by WLAN A.}    
			\label{fig:central_wlan_throughput}
		\end{subfigure}
		
		\caption{Throughput analysis of the \textit{central WLAN scenario.}}
		\label{fig:results_central}
	\end{figure}

	Figure \ref{fig:probability_throughput_load_similar} shows the probability of WLAN A to successfully transmit its traffic load, i.e., $P_\text{A} = \mathcal{P}\big(\Gamma_\text{A} \geq (1 - \epsilon_\Gamma) \ell_\text{A}\big)$. Note that we use a margin of error $\epsilon_\Gamma = 0.05$ to cope with the stochastic packet generation of the performed simulations. \newver{The average throughput of A for each of the policies is plotted in Figure \ref{fig:central_wlan_throughput}}.
	As expected, SCB is viable only for few scenarios when the traffic load is extremely small. \newver{This is because the rest of WLANs most likely prevent A to initiate any transmission by occupying part of its allocated channel $C_\text{A}$.}
	Instead, the other policies perform much better -- specially, AM -- since they avoid saturation with high probability even for high traffic loads. While A avoids saturation in some scenarios for $\ell_\text{A} < 92.16$ Mbps with OP and $\ell_\text{A} < 122.88$ Mbps with PU, respectively, the aggressive adaptability nature of AM allows avoiding saturation in scenarios even where $\ell_\text{A} = 184.32$ Mbps.
	\newver{Nevertheless, we note that there are scenarios where AM heavily suffers from the hidden node problem since the SINR at the receiver is importantly reduced. Accordingly, even though the AP may find the whole spectrum free, the STA is not able to decode most of the RTS packets due to interference. This effect can be clearly seen in the $P_\text{A}$ improvement of OP or PU for low traffic loads (i.e., $<$ 46.08 Mbps).}
	
\begin{figure}[t]
		\centering
	
		\begin{subfigure}[b]{0.475\textwidth}  
			\centering 
			\includegraphics[width=\textwidth]{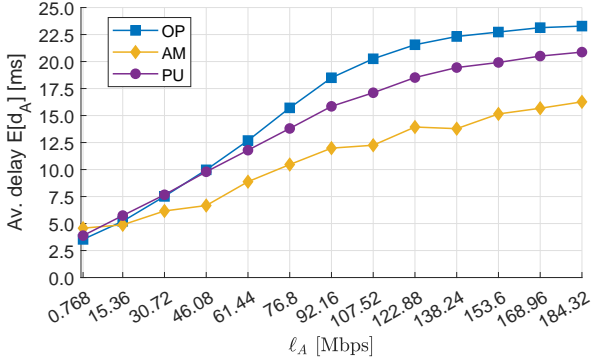}
			\caption[]{Traffic load effect on the average delay \newver{(without outliers)} experienced by WLAN A.}
			\label{fig:central_wlan_delay}
		\end{subfigure}
		~
		\begin{subfigure}[b]{0.475\textwidth}  
			\centering 
			\includegraphics[width=\textwidth]{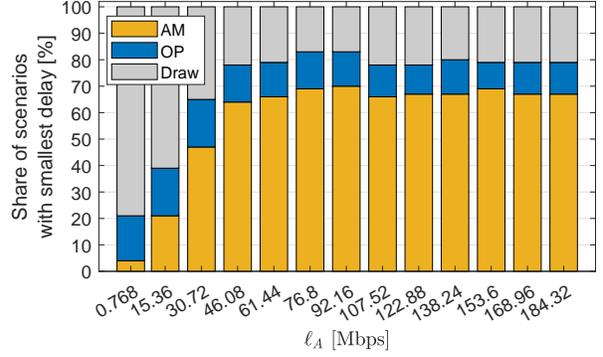}
			\caption[]{Share of scenarios where AM or OP provides the smallest delay for A.}
			\label{fig:central_wlan_share_op}
		\end{subfigure}
		
		\begin{subfigure}[b]{0.475\textwidth}  
			\centering 
			\includegraphics[width=\textwidth]{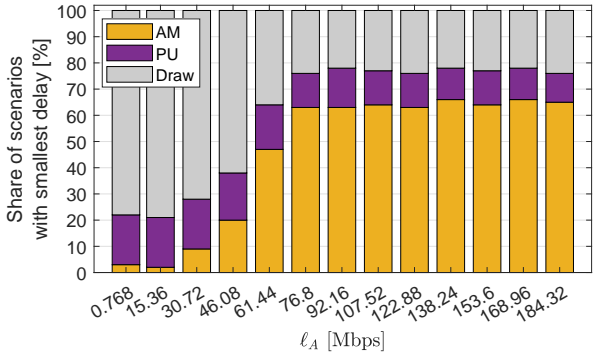}
			\caption[]{Share of scenarios where AM or PU provides the smallest delay for A.}
			\label{fig:central_wlan_share}
		\end{subfigure}
		~
		\begin{subfigure}[b]{0.475\textwidth}  
			\centering 
			\includegraphics[width=\textwidth]{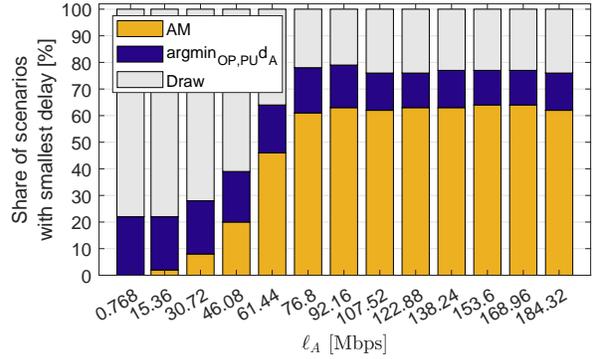}
			\caption[]{Share of scenarios where AM or the best policy among OP and PU provides the smallest delay for A.}
			\label{fig:central_wlan_share_am_opORpu}
		\end{subfigure}
		
		\caption{Delay analysis of the \textit{central WLAN scenario}.}
		\label{fig:results_central_2}
	\end{figure}

	The average packet delay experienced by A under different traffic loads is shown in Figure \ref{fig:central_wlan_delay}. \newver{For the sake of representation, we consider only those scenarios free of outliers.\footnote{\newver{A scenario is labeled as outlier if $d_\text{A} \geq 100$ \si{ms}. Only 3\% of the scenarios where $\mathcal{D}_\text{A} = \text{OP},\text{AM},\text{PU}$ are outliers.}}. Note that we do not plot the delay of SCB since its performance is clearly deficient, as shown by Figures \ref{fig:probability_throughput_load_similar} and \ref{fig:central_wlan_throughput}.} As a significant result, we note that the smallest average delay is provided by AM for all the studied loads\newver{, except for $\ell_\text{A} = 0.768$ Mbps. This proves that the delay reduction observed by OP at low traffic loads does also hold for uncontrolled scenarios where neighboring WLANs may have different CB policies and higher traffic loads.}
	
	
	\newver{Despite the average superior performance of AM, there are few scenarios where a less aggressive approach like PU or even OP outperforms it.}
	In this regard, we assess below the share of scenarios where each policy provides the smallest average packet delay for WLAN A. In particular, Figure \ref{fig:central_wlan_share_op} compares AM against OP and Figure \ref{fig:central_wlan_share} does so for AM and PU. \newver{Finally, Figure \ref{fig:central_wlan_share_am_opORpu} compares the number of scenarios where AM is better than the best combination of OP and PU, i.e., AM is compared against $\mathcal{D}=\argmin_\text{{OP,PU}} d_A$ for each of the simulated scenario}. \newver{We say that for any given pair of policies $\mathcal{D}_1$, $\mathcal{D}_2$, three types of outcomes are categorized according to a predefined delay margin $\delta_d = 1$ ms:}
	\begin{equation*}
		\texttt{if }\text{E}\big[d_\text{A}^{\mathcal{D}_1}\big] - \text{E}\big[d_\text{A}^{\mathcal{D}_2}\big]
		\begin{cases}
		< -\delta_d, & \mathcal{D}_1\text{ better than } \mathcal{D}_2\\
		> \delta_d, & \mathcal{D}_2\text{ better than } \mathcal{D}_1\\
		\text{otherwise}, & \text{ draw}
		\end{cases} \text{.}
	\end{equation*}
	\newver{The delay margin allows us capturing those cases where $\mathcal{D}_1$ and $\mathcal{D}_2$ perform similarly.}
	
	We see that in most of the cases AM outperforms OP and PU, specially for scenarios with mid-high traffic loads.
	\newver{Nonetheless, for low loads, we note that there is always a better choice than AM for reducing the delay, corroborating the outcomes from previous sections. In  addition,}
	there is a significant share of scenarios where OP and, especially, PU provide similar or \newver{even} smaller delays than AM for all traffic loads.
	This mainly occurs when A and its neighboring nodes are able to concurrently transmit in different channels through interactions that are not given when implementing AM.
	Essentially, when A transmits in its whole available bandwidth, neighboring WLANs with primary channels overlapping with A's transmission must wait until it is finished. Afterwards, such WLANs are able to terminate their backoffs and could select a transmission channel including A's primary in turn. This generates \textit{all or nothing} states like the one shown in \textit{Scenario I} that keep A's backoff frozen for longer periods of time. Instead, if A transmits in narrower channels by implementing OP or PU, such WLANs could transmit at the same time in non-overlapping channels and enable more successful parallel transmissions.
	
	In summary, we see that overlapping approaches can significantly enhance traditional single-channel performance in terms of delay and throughput in uncontrolled and realistic HD deployments. Still, there are cases when an overlapping approach that always selects the maximum available bandwidth can be counterproductive in the mid/long-term. 
	Despite the intrinsic uncertainty of spatially distributed WLAN deployments, we can state as a rule of thumb that DCB is convenient when applied through spectrum-adapting policies. Nonetheless, as indicated by the scenarios where OP and/or PU outperformed AM, there is room for further improvement through smarter adaptation by adopting policies on a per-WLAN basis.
	\newver{Hence}, we envision that the most effective way of using DCB is to allocate all the nodes with the whole \newver{available} frequency spectrum, and \newver{to} smartly assign the primary channel. Moreover, deeper improvements could be achieved by endowing the nodes with \newver{the} capability to recover from lousy situations like FIM, which are more likely to happen when neighboring WLANs implement aggressive DCB.
	
	\section{Conclusions} \label{sec:conclusions}
	
	In this work\newver{,} we assess the performance of CB in WLANs under different traffic loads. By modeling and simulating CB policies in spatially distributed scenarios we shed light on the question: is it convenient to share wider channels and potentially overlap in \newver{the} spectrum or not?
	We show that, while the performance of SCB is clearly poor for moderate-high traffic loads, spectrum-adapting DCB can significantly outperform the traditional single-channel approach in terms of throughput and delay, even in high-density deployments. \newver{However, we also remark two main outcomes concerning DCB: \textit{i}) the intrinsic risk when it comes to generating unfair scenarios as a consequence of the hidden node and FIM situations, and \textit{ii}) the exposure to doing more harm than good in terms of delay for low traffic regimes.}
	
	In this regard, the intricate nature of uncontrolled WLAN deployments leaves room for further improvements \newver{in spectrum efficiency, while prevents designing effective predefined rules. Therefore, our next work will focus on two aspects: studying machine learning based DCB policies to efficiently adapt to traffic load needs, and jointly combining DCB with adequate primary channel allocation.}
	
	
	\section*{Acknowledgment}
	
	This work has been partially supported by a Gift from CISCO University Research Program (CG\#890107) \& Silicon Valley Community Foundation, by the Spanish Ministry of Economy and Competitiveness under the Maria de Maeztu Units of Excellence Programme (MDM-2015-0502), and by the Catalan Government under grant SGR-2017-1188. The work by S. Barrachina-Mu\~noz is supported by an FI grant from the Generalitat de Catalunya.

	%

	\bibliographystyle{unsrt}
	\bibliography{bib}

\begin{thebibliography}{10}

\bibitem{elbamby2018toward}
M.~Elbamby, C.~Perfecto, M.~Bennis, and K.~Doppler.
\newblock Toward low-latency and ultra-reliable virtual reality.
\newblock {\em IEEE Network}, 32(2):78--84, 2018.

\bibitem{80211n}
{IEEE 802.11n. Standard for Wireless LAN Medium Access Control (MAC) and
  Physical Layer (PHY): Enhancements for High Throughput}.
\newblock {\em IEEE}, 2009.

\bibitem{80211ac}
I.~P802.11ac.
\newblock {Standard for Wireless LAN Medium Access Control (MAC) and Physical
  Layer (PHY) specifications: Enhancements for Very High Throughput for
  Operation in Bands below 6 GHz.}
\newblock {\em IEEE}, 2014.

\bibitem{802.11ax}
{IEEE 802.11 Task Group AX. Status of Project IEEE 802.11ax High Efficiency
  WLAN (HEW)}.
\newblock \url{http://www.ieee802.org/11/Reports/tgax_update.htm}.
\newblock Accessed: 2018-09-21.

\bibitem{80211ET}
{EXtreme Throughput (XT) 802.11}.
\newblock {\em Doc.: IEEE 802.11-18/0789r10}, 2018.

\bibitem{bukhari2016survey}
S.~Bukhari, M.~Rehmani, and S.~Siraj.
\newblock {A survey of channel bonding for wireless networks and guidelines of
  channel bonding for futuristic cognitive radio sensor networks}.
\newblock {\em IEEE Communications Surveys \& Tutorials}, 18(2):924--948, 2016.

\bibitem{park2011ieee}
M.~Park.
\newblock {IEEE 802.11 ac: Dynamic bandwidth channel access}.
\newblock In {\em Communications (ICC), 2011 IEEE International Conference on},
  pages 1--5. IEEE, 2011.

\bibitem{deek2011impact}
L.~Deek, E.~Garcia-Villegas, E.~Belding, S.~Lee, and K.~Almeroth.
\newblock The impact of channel bonding on 802.11 n network management.
\newblock In {\em Proceedings of the Seventh Conference on emerging Networking
  Experiments and Technologies}, page~11. ACM, 2011.

\bibitem{arslan2010auto}
M.~Arslan, K.~Pelechrinis, I.~Broustis, S.~Krishnamurthy, S.~Addepalli, and
  K.~Papagiannaki.
\newblock Auto-configuration of 802.11n wlans.
\newblock In {\em Proceedings of the 6th International Conference}, page~27.
  ACM, 2010.

\bibitem{arslan2013acorn}
M.~Arslan, K.~Pelechrinis, I.~Broustis, S.~Krishnamurthy, S.~Addepalli, and
  K.~Papagiannaki.
\newblock {ACORN: An auto-configuration framework for 802.11 n WLANs}.
\newblock {\em IEEE/ACM Transactions on Networking (TON)}, 21(3):896--909,
  2013.

\bibitem{shrivastava2008802}
V.~Shrivastava, S.~Rayanchu, J.~Yoonj, and S.~Banerjee.
\newblock 802.11n under the microscope.
\newblock In {\em Proceedings of the 8th ACM SIGCOMM conference on Internet
  measurement}, pages 105--110. ACM, 2008.

\bibitem{bellalta2016ieee}
B.~Bellalta.
\newblock {IEEE} 802.11ax: High-efficiency {WLANs}.
\newblock {\em IEEE Wireless Communications}, 23(1):38--46, 2016.

\bibitem{deek2014intelligent}
L.~Deek, E.~Garcia-Villegas, E.~Belding, S.~Lee, and K.~Almeroth.
\newblock Intelligent channel bonding in 802.11 n wlans.
\newblock {\em IEEE Transactions on Mobile Computing}, 13(6):1242--1255, 2014.

\bibitem{deek2013joint}
L.~Deek, E.~Garcia-Villegas, E.~Belding, S.~Lee, and K.~Almeroth.
\newblock {Joint rate and channel width adaptation for 802.11 MIMO wireless
  networks}.
\newblock In {\em Sensor, Mesh and Ad Hoc Communications and Networks (SECON),
  2013 10th Annual IEEE Communications Society Conference on}, pages 167--175.
  IEEE, 2013.

\bibitem{wang2016managing}
W.~Wang, F.~Zhang, and Q.~Zhang.
\newblock Managing channel bonding with clear channel assessment in 802.11
  networks.
\newblock In {\em Communications (ICC), 2016 IEEE International Conference on},
  pages 1--6. IEEE, 2016.

\bibitem{gong2011channel}
M.~Gong, B.~Hart, L.~Xia, and R.~Want.
\newblock {Channel bounding and MAC protection mechanisms for 802.11ac}.
\newblock In {\em Global Telecommunications Conference (GLOBECOM 2011), 2011
  IEEE}, pages 1--5. IEEE, 2011.

\bibitem{barrachina2018performance}
S.~Barrachina-Mu{\~n}oz, F.~Wilhelmi, and B.~Bellalta.
\newblock {Dynamic Channel Bonding in Spatially Distributed High-Density
  WLANs}.
\newblock {\em arXiv preprint arXiv:1801.00594}, 2018.

\bibitem{zeng2014first}
Y.~Zeng, P.~Pathak, and P.~Mohapatra.
\newblock A first look at 802.11ac in action: Energy efficiency and
  interference characterization.
\newblock In {\em Networking Conference, 2014 IFIP}, pages 1--9. IEEE, 2014.

\bibitem{yun2013fine}
S.~Yun, Daehyeok Kim, and L.~Qiu.
\newblock Fine-grained spectrum adaptation in wifi networks.
\newblock In {\em Proceedings of the 19th annual international conference on
  Mobile computing \& networking}, pages 327--338. ACM, 2013.

\bibitem{wang2017wideband}
W.~Wang, Y.~Chen, Z.~Wang, J.~Zhang, K.~Wu, and Q.~Zhang.
\newblock Wideband spectrum adaptation without coordination.
\newblock {\em IEEE Transactions on Mobile Computing}, 16(1):243--256, 2017.

\bibitem{huang2016dynamic}
P.~Huang, X.~Yang, and L.~Xiao.
\newblock {Dynamic channel bonding: enabling flexible spectrum aggregation}.
\newblock {\em IEEE Transactions on Mobile Computing}, 15(12):3042--3056, 2016.

\bibitem{nabil2017adaptive}
A.~Nabil, M.~Abdel-Rahman, and A.~MacKenzie.
\newblock {Adaptive Channel Bonding in Wireless LANs Under Demand Uncertainty}.
\newblock {To appear in the Proceedings of the IEEE International Symposium on
  Personal, Indoor and Mobile Radio Communications (PIMRC), Montreal, QC,
  Canada, October 2017.}

\bibitem{chen2018dbs}
Y.~Chen, D.~Wu, T.~Sung, and K.~Shih.
\newblock {DBS: A dynamic bandwidth selection MAC protocol for channel bonding
  in IEEE 802.11 ac WLANs}.
\newblock In {\em Wireless Communications and Networking Conference (WCNC),
  2018 IEEE}, pages 1--6. IEEE, 2018.

\bibitem{khairy2018renewal}
S.~Khairy, M.~Han, L.~Cai, Y.~Cheng, and Z.~Han.
\newblock {A Renewal Theory based Analytical Model for Multi-channel Random
  Access in IEEE 802.11ac/ax}.
\newblock {\em IEEE Transactions on Mobile Computing}, 2018.

\bibitem{byeon2018reconn}
S.~Byeon, H.~Kwon, Y.~Son, C.~Yang, and S.~Choi.
\newblock {RECONN: Receiver-Driven Operating Channel Width Adaptation in IEEE
  802.11ac WLANs}.
\newblock In {\em IEEE INFOCOM 2018-IEEE Conference on Computer
  Communications}, pages 1655--1663. IEEE, 2018.

\bibitem{bellalta2014channel}
B.~Bellalta, A.~Faridi, J.~Barcelo, A.~Checco, and P.~Chatzimisios.
\newblock {Channel bonding in short-range WLANs}.
\newblock In {\em European Wireless 2014; 20th European Wireless Conference;
  Proceedings of}, pages 1--7. VDE, 2014.

\bibitem{bellalta2016interactions}
B.~Bellalta, A.~Checco, A.~Zocca, and J.~Barcelo.
\newblock {On the interactions between multiple overlapping WLANs using channel
  bonding}.
\newblock {\em IEEE Transactions on Vehicular Technology}, 65(2):796--812,
  2016.

\bibitem{han2016performance}
M.~Han, Sami K., L.~X. Cai, and Y.~Cheng.
\newblock {Performance Analysis of Opportunistic Channel Bonding in
  Multi-Channel WLANs}.
\newblock In {\em Global Communications Conference (GLOBECOM), 2016 IEEE},
  pages 1--6. IEEE, 2016.

\bibitem{kim2017throughput}
M.~Kim, T.~Ropitault, S.~Lee, and N.~Golmie.
\newblock {A Throughput Study for Channel Bonding in IEEE 802.11ac Networks}.
\newblock {\em IEEE Communications Letters}, 2017.

\bibitem{kai2017channel}
C.~Kai, Y.~Liang, T.~Huang, and X.~Chen.
\newblock {A Channel Allocation Algorithm to Maximize Aggregate Throughputs in
  DCB WLANs}.
\newblock {\em arXiv preprint arXiv:1703.03909}, 2017.

\bibitem{barrachina2018komondor}
S.~Barrachina-Mu{\~n}oz, F.~Wilhelmi, I.~Selinis, and B.~Bellalta.
\newblock {Komondor: a Wireless Network Simulator for Next-Generation
  High-Density WLANs}.
\newblock {\em arXiv preprint arXiv:1811.12397}, 2018.

\bibitem{laufer2013capacity}
R.~Laufer and L.~Kleinrock.
\newblock {On the capacity of wireless CSMA/CA multihop networks}.
\newblock In {\em INFOCOM, 2013 Proceedings IEEE}, pages 1312--1320. IEEE,
  2013.

\bibitem{faridi2016analysis}
A.~Faridi, B.~Bellalta, and A.~Checco.
\newblock {Analysis of Dynamic Channel Bonding in Dense Networks of WLANs}.
\newblock {\em IEEE Transactions on Mobile Computing}, 2016.

\bibitem{chatzimisios2004performance}
P.~Chatzimisios, A.~Boucouvalas, and V.~Vitsas.
\newblock {Performance analysis of IEEE 802.11 DCF in presence of transmission
  errors}.
\newblock In {\em Communications, 2004 IEEE International Conference on},
  volume~7, pages 3854--3858. IEEE, 2004.

\bibitem{xu2007indoor}
D.~Xu, J.~Zhang, X.~Gao, P.~Zhang, and Y.~Wu.
\newblock {Indoor office propagation measurements and path loss models at 5.25
  GHz}.
\newblock In {\em Vehicular Technology Conference, 2007. VTC-2007 Fall. 2007
  IEEE 66th}, pages 844--848. IEEE, 2007.

\end{thebibliography}
	
	\appendix
	
	\section{Evaluation setup}
	
	\begin{table}[t]
		\caption{Evaluation setup (from simulation setup in \cite{barrachina2018performance}).}
		\label{table:appendix_table}
		\centering
		\footnotesize
		\begin{tabularx}{.7\textwidth}{ccc}
			\toprule
			\textbf{Parameter}     & \textbf{Description}              & \textbf{Value} \\ 
			\midrule
			$f_\text{c}$ & Central frequency           & 5.25 GHz  \\
			$\abs c$ & Basic channel bandwidth          & 20 MHz \\
			$L_\text{D}$       & Data packet size           & 12000 bits     \\ 
			$N_\text{b}$		& Buffer capacity & 150 packets \\
			$N_\text{a}$       & Max. no. of packets in a frame & 64             \\  
			$\text{CW}_\text{min}$ & Min. contention window            & 16             \\ 
			$m$                    & No. of backoff stages          & 5              \\   
			MCS						& IEEE 802.11ax MCS index							& 0 - 11		\\
			$\eta$                 & MCS's packet error rate        & 0.1           \\ 
			CCA                    & CCA threshold                               & -82 dBm        \\ 
			$P_\text{tx}$          & Transmission power                & 15 dBm         \\ 
			$G_\text{tx}$         & Transmitting gain                 & 0 dB           \\ 
			$G_\text{rx}$         & Reception gain                    & 0 dB           \\ 
			$\text{PL}(d)$		& Path loss 			& see (\ref{eq:path_loss})		\\
			\newver{$\text{P}_\nu$}                    & \newver{Adjacent power leakage factor}          & \newver{-20 dB}          \\ 
			CE                     & Capture effect threshold          & 20 dB          \\ 
			$N$                      & Background noise level            & -95 dBm        \\
			\midrule
			$T_\text{e}$       & Empty backoff slot duration                     & 9 $\mu$s          \\
			$T_\text{SIFS}$                   & SIFS duration                     & 16 $\mu$s      \\ 
			$T_\text{DIFS}$                   & DIFS duration                     & 34 $\mu$s      \\ 
			$T_\text{PIFS}$                   & PIFS duration                     & 25 $\mu$s      \\
			$T_\text{PHY-leg}$      & Legacy preamble     & 20 $\mu$s           \\ 
			$T_\text{PHY-HE-SU}$      & HE single-user preamble       & 164 $\mu$s \\
			$\sigma_\text{leg}$      & Legacy OFDM symbol duration     & 4 $\mu$s           \\
			$\sigma$      & OFDM symbol duration     & 16 $\mu$s           \\
			$L_\text{BACK}$       & Length of a block ACK             & 432 bits       \\ 
			$L_\text{RTS}$        & Length of an RTS packet           & 160 bits       \\ 
			$L_\text{CTS}$        & Length of a CTS packet            & 112 bits       \\ 
			$L_\text{SF}$      & Length of service field       & 16 bits           \\ 
			$L_\text{MD}$      & Length of MPDU delimiter       & 32 bits           \\ 
			$L_\text{MH}$      & Length of MAC header     & 320 bits           \\ 
			$L_\text{TB}$      & Length of tail bits     & 18 bits           \\ 
			\bottomrule
			
		\end{tabularx}
	\end{table}
	
	The values of the parameters considered in the simulations presented throughout this paper are shown in Table \ref{table:appendix_table}.
	Regarding the path loss, we use the dual-slope log-distance model for 5.25 GHz indoor environments in room-corridor condition \cite{xu2007indoor}. Specifically, the path loss in dB experienced at a distance $d$ is defined by
	\begin{equation}	\label{eq:path_loss}
	\text{PL}(d) = 
	\left\{
	\begin{array}{ll}
	53.2 \, \Plus \, 25.8 \log_{10}(d)  & \mbox{if } d \leq d_1 \text{ m} \\
	56.4 \, \Plus \, 29.1 \log_{10}(d)  & \mbox{otherwise}
	\end{array}
	\right.\text{,}
	\end{equation} 
	where $d_1 = 9$ m is the break point distance.
	
	The MCS index used for each possible channel bandwidth (i.e., 20, 40, 80 or 160 MHz) was the highest allowed according to the power power budget established between the WLANs and their corresponding STA/s and the minimum sensitivity required by the MCSs. As stated by the IEEE 802.11ax amendment, the number of transmitted bits per OFDM symbol used in the data transmissions is given by the channel bandwidth and the MCS parameters, i.e., $r = Y_\text{sc} Y_\text{m} Y_\text{c} V_\text{s}$, where $Y_\text{sc}$ is the number of data sub-carriers, $Y_\text{m}$ is the number of bits in a modulation symbol, $Y_\text{c}$ is the coding rate, and $ V_\text{s} = 1$ is the number of single user spatial streams (note that we only consider one stream per transmission).
	
	The number of data sub-carriers depends on the transmission channel bandwidth. Specifically, $Y_\text{sc}$ can be 234, 468, 980 or 1960 for 20, 40, 80, and 160 MHz, respectively. For instance, the data rate provided by MCS 11 in a 20 MHz transmission is $s = (234 \times 10 \times 5/6 \times 1)\sigma^{-1} = 121.9$ Mbps.
	However, control frames are transmitted in legacy mode using the basic rate $r_\text{leg} = 24$ bits per OFDM symbol of MCS 0, corresponding to $s_\text{leg} = 6$ Mbps since the legacy OFDM symbol duration $\sigma_\text{leg}$ must be considered. With such parameters we can compute the duration of each packet transmission:
	\begin{flalign*}
	T_\text{RTS} &= T_\text{PHY-leg} \Plus \ceil*{\frac{L_\text{SF} \Plus L_\text{RTS} \Plus L_\text{TB}}{r_\text{leg}}} \sigma_\text{leg}  \text{,} \\
	T_\text{CTS} &= T_\text{PHY-leg} \Plus \ceil*{\frac{L_\text{SF} \Plus L_\text{CTS} \Plus L_\text{TB}}{r_\text{leg}}} \sigma_\text{leg} \text{,} \\
	T_\text{DATA}(N_a) &= T_\text{PHY-HE-SU} \Plus \ceil*{\frac{L_\text{SF} \Plus N_\text{a} (L_\text{MD} \Plus L_\text{MH} \Plus L_\text{D})  \Plus L_\text{TB}}{r}} \sigma \text{,} \\
	T_\text{BACK} &= T_\text{PHY-leg} \Plus \ceil*{\frac{L_\text{SF} \Plus L_\text{BACK} \Plus L_\text{TB}}{r_\text{leg}}} \sigma_\text{leg} \text{.}
	\end{flalign*}

\end{document}